**The Scientific and Historical Value of Annotations on Astronomical Photographic Plates**

Sara J. Schechner

Harvard University

David Sliski

University of Pennsylvania

**Abstract**

The application of photography to astronomy was a critical step in the development of astrophysics at the end of the nineteenth century.  Using custom-built photographic telescopes and objective prisms, astronomers took images of the sky on glass plates during a 100-year period from many observing stations around the globe.  After each plate was developed, astronomers and their assistants studied and annotated the plates as they made astrometric, photometric and spectroscopic measurements, counted galaxies, observed stellar variability, tracked meteors, and calculated the ephemerides of asteroids and comets.  In this paper, the authors assess the importance of the plate annotations for future scientific, historical, and educational programs.  Unfortunately, many of these interesting annotations are now being erased when grime is removed from the plates before they are digitized to make the photometric data available for time-domain astrophysics.  To see what professional astronomers and historians think about this situation, the authors conducted a survey.  This paper captures the lively discussion on the pros and cons of the removal of plate markings, how best to document them if they must be cleaned off, and what to do with plates whose annotations are deemed too valuable to be erased. Three appendices to the paper offer professional guidance on the best practices for handling and cleaning the plates, photographing any annotations, and rehousing them.

The Digital Access to a Sky Century at Harvard (DASCH), a multiyear undertaking of the Harvard College Observatory, is digitizing the world's largest collection of glass-plate photographs captured of the sky between 1885 and 1993 with observing stations in Massachusetts, Peru, South Africa, and elsewhere around the globe. The purpose of the ambitious project involving more than 500,000 plates is to make the rich trove of data held in the Harvard Astronomical Plate Stacks accessible for time-domain astrophysics. About 20% of these plates have historically significant markings related to the work of the Harvard College Observatory on topics such as stellar variability, proper motion, galaxy dynamics, cosmology, meteors, and the ephemerides of minor planets, asteroids, and comets. Names associated with the research performed with the plates include Edward Pickering, Williamina Fleming, Annie Jump Cannon, Henrietta Leavitt, Harlow Shapley, Fred Whipple, and others. Historically significant writing is also found on the plate jackets.[1] (See Figs. 1-2)

In May 2013 Harvard professor Jonathan Grindlay, Robert Treat Paine Professor of Practical Astronomy and the lead scientist of DASCH, called a meeting to discuss the preservation of the photographic glass plates that were being scanned.  Present at this meeting were Harvard museum curators, librarians, photographic conservators, historians of science, scientists, and the DASCH staff.[2]

One of the major issues discussed without conclusion was what to do with the annotations marked on the non-emulsion side of glass plates. The scientists argued that grime and annotations obscured the scientific data (i.e., the positions and brightness of celestial objects) that they hoped to preserve and make accessible by creating a digital file of each astronomical photograph. Therefore, project protocols called for a two-step process.  A technician took a photograph of the annotations on the plate and its jacket.  Then each plate was cleaned with razor blades, an ethanol/water (40/60) solution, and microfiber cloths in order to remove smudges and India ink annotations from the non-emulsion side before the plate was scanned.  The historians, librarians, and conservators were worried that the solvents applied by hand on each plate and then wiped off with towels, and those to be employed with stainless-steel wire brushes in an automated plate washing machine (still in the works) might accidentally damage the emulsion or scratch the glass.[3]  They were also concerned that the cleaning of the plates before scanning was an irreversible process whereby valuable historical and scientific data possibly contained in the annotations would be lost.  They asked about the quality and format of the photographs of annotations, and made recommendations to increase resolution, optimize lighting, calibrate color, change file type, and improve the photography station.  After the meeting, the conservators and photographer submitted reports on best practices for plate handling and photography to DASCH.[4]  Updated versions of the reports are appended to this paper for the benefit of others working with photographic plates.

The 2013 meeting, prompted specific questions:

1. To what degree do the annotations on the plates and their paper jackets have historical value?

2. How should they be documented?

3. Are photographs of the annotated plates and jackets sufficient substitutes for the real artifacts for research?   If yes, what resolution and image quality would satisfy researchers?

4.  Should the original plates and jackets be preserved after digitization?

5.  Should any annotated plates be set aside uncleaned in order to illustrate the work of specific researchers, their methods, research subjects, and major discoveries?  If so, which ones and how many?

In order to answer these questions, Professor Jonathan Grindlay, asked Dr. Sara Schechner, curator of Harvard's Collection of Historical Scientific Instruments, past chair of the Historical Astronomy Division (HAD) of the American Astronomical Society (AAS), and a founding member of the AAS Working Group on the Preservation of Astronomical Heritage (WGPAH), to conduct a survey of the history-of-astronomy community. Schechner invited David Sliski, then a DASCH curatorial assistant, to collaborate, since he had raised concerns about plate handling, organized the advisory meeting, and helped to implement many of the changes suggested by museum professionals. This report summarizes the findings of that survey. Illustrations of annotated plates and their jackets are included in order to assist readers in understanding what such plates look like.

**Method of the Survey**

The survey was qualitative with 25 people responding in writing to a questionnaire that was distributed via email to WGPAH, HAD, HASTRO-L (a history of astronomy listserv), and personal correspondents. Prior to the survey, members of WGPAH and HAD were less likely to be strangers to the challenges of preserving astronomical photographs and archival records, since these organizations had been instrumental in organizing meetings and publications on the topic.[5] Subscribers to HASTRO-L, on the other hand, are drawn from a wider population than HAD's and WGPAH's professional astronomers, and they would have been less familiar with the survey topic. Eighty percent (80%) of the respondents were astronomers, many of whom had worked with photographic plates in their research. The remaining twenty percent (20%) were professional historians of science.

Although the response rate was small for a survey circulated to several hundred individuals, the respondents were distinguished scientists and historians. Most replies were public insofar as they were sent to discussion lists where all subscribers could read them and openly engage with them. The result was a lively discussion, which this report strives to capture in the fashion of oral histories by having many quotations.

**Summary of Findings**

All respondents agreed that the annotations on the plates and jackets had scientific, historical, and educational value, and the preponderance rated this value as very high. Most believed that preservation of this scientific and historical value could be accomplished adequately by photography, provided that the images were of very high quality. All agreed that some plates should be set aside without removing the annotations as samples of the historical methods, the work of key individuals, and important discoveries.

<div align="center">

**Discussion of Particular Questions**

</div>

***Do the plate markings and their jackets have any historical or scientific value?***

The answer was a resounding yes, and respondents stated many reasons why. The best digest was offered by David DeVorkin, Senior Curator at the Smithsonian National Air and Space Museum: The annotations, he said, were "incontrovertible proof that something was done in a certain way, without the possibility of some equivalent of 'Photoshopping,' either conscious or inadvertent".[6]

Only one respondent, Bradley Schaefer, who described himself as "both a heavy user of the Harvard plates (and other plate collections) and a person who has done a lot of history work", claimed that he had "never come across any annotation that was of any astrophysical…[or] historical use at all". But after reading the discussion on HASTRO-L, he agreed that other researchers have found utility, writing "*I* have never found the plate annotations helpful, but Wayne Osborn's email trumps my lack of utility, because *he* has found a variety of uses for them".[7]  While acknowledging that some historical and representative plate annotations should be saved, Schaefer maintained that most annotations would never be of utility to astronomers or historians.  (Wayne Osborn of Yerkes Observatory, to whom Schaefer referred, has been a scientific user of astronomical photographic plates and a major advocate for their preservation.[8])

      Schaefer is to be credited with voicing an opinion held strongly by others, but there were also many respondents at the opposite pole. The latter believed so intensely in the value of the markings that they contended that any removal of these would be detrimental to future research, both historical and scientific.  Arguing for the historical importance, Barry Madore, a senior researcher at the Observatories of the Carnegie Institution for Science, said that "unless a strong scientific case can be made for scrubbing any given plate, the over-riding default should be to preserve the historical record. There is probably more value in the history than in the developed grains". Wendy Freedman, now at the University of Chicago, but then director of the Carnegie Observatories, concurred "that removing markings is to be avoided given the historical nature of these markings". Jay Pasachoff, director of the Hopkins Observatory at Williams College, wrote, "Just as I wouldn't scrub off Edwin Hubble's handwriting, I would say that the handwriting of Henrietta Leavitt, Williamina Fleming, and others shouldn't be scrubbed off", but added that they should "at least be separately scanned".[9]  (See Fig. 3 for a plate related to Leavitt's work.)

Arguing on behalf of the scientific importance of markings were astronomers such as William Liller, R. W. Willson Professor of Applied Astronomy at Harvard (1961-1983), who had spent many hours in the Harvard plate stacks in the 1970s identifying x-ray sources and studying their historical behavior, as well as noting quasi-stellar objects, novae, supernovae, and the occasional asteroid or comet of interest. Liller made this "fervent plea":

> I herewith submit that in the vast majority of cases, the ever-so-tiny little "vee" marks or circles on the glass plates were put there in indelible ink by highly diligent inspectors who had a clear understanding that at some time in the future, someone might just want to measure precisely the position or brightness of the object of interest. Ergo, meticulous care was almost certainly taken not to let the ink mark encroach on, or even come close to the image. And so my strong recommendation would be not to "scrub off" these marks.[10]

(See Fig. 4 for an example of such plate markings.)

Even those who took the opposite stance—i.e., that all plates should be cleaned—still conceded that the markings might prove worthwhile to document. Vladimir Strelnitski, the retiring director of Maria Mitchell Observatory on Nantucket, recalled discussions on the matter at the international meeting at the Pisgah Astronomical Research Institute (PARI) in Rosman, North Carolina in November 2007.[11] "Some people expressed a strong belief that the old markings on the plates may be useful, and thus they should be copied before cleaning the plate. It was too late an advice for MMO:  the scanning had been finished by then". His frank opinion was that the old markings "may present some interest (mostly historical) only in very rare cases".  Therefore, their preservation should be decided case by case in big scanning projects where there was pressure to get things done as quickly as possible.[12]  Strelnitski's colleague at the Maria Mitchell Association, and formerly the Executive Officer for the AAS (1979-1995), Peter Boyce only half agreed: "I would not save any of the plate markings—as long as they are photographed sufficiently well".[13]

Indeed some respondents expressed curiosity and dismay about missed opportunities arising from cases of "lost" annotations.  Virginia Trimble, an astrophysicist and historian of astronomy at the University of California Irvine, and Lee Robbins, Head Librarian of the Astronomy and Astrophysics Library of the University of Toronto and co-author with Wayne Osborn of a plate census and reports on plate preservation, both shared the story of Harlow Shapley and Milton Humason.  Humason's connections to astronomy started as a mule driver hauling supplies up Mount Wilson for the construction of the 100-inch telescope.  When it was complete in 1917, he was hired as a janitor and a couple of years later became a member of the scientific staff.  Shapley was on the mountain at the same time, and offered to let the budding astronomer have a look at his M31 plates with a blink comparator.  After a few weeks of studying the plates, Humason returned with several objects marked and asked if they might be Cepheid variables.  According to Humason, Shapley said he must be mistaken; everyone knows those variables cannot be Cepheids because the Andromeda nebula is part of our galaxy and any Cepheid in it would be brighter.  He then took out his handkerchief and rubbed out Humason's marks.  If the story is credible—as Shapley later admitted—Shapley erased an opportunity for recognizing the extragalactic nature of the spiral nebula four to five years before Hubble did.  He also erased our chance to check Humason's story.[14]

### What Kind of Research Might Be Done from Annotations in the Future?

Respondents envisioned many general scientific and historical reasons to consult the annotations of photographic plates and their jackets.  In addition to the obvious point that the jackets record essential information regarding exposure dates, times, equatorial position of the plate center, and plate number, the principal reasons were:

- As proof of whom had taken, read, or examined the plate.

- To confirm a result.

- To disclose whether reference stars or other objects were misidentified.

- To see if markings somehow influenced or corrupted the data.

- To understand what choices were made that led to a particular result.

- To recognize a first discovery.

- To derive cultural patterns and practices in observing.

The annotations were often compared to book marginalia insofar as they might help scholars to understand whom had read what, why they had read it, and what their conclusions were.[15] One surprise of the survey was that most reasons given *not* to erase plate markings were scientific rather than historical.  Although all of the stated reasons have great potential, historical value, only the last item on the list above—to derive cultural patterns and practices in observing—is exclusively a historical reason.[16]

*Have plate markings been useful for research and education in the past?*

The DASCH principal investigator, Jonathan Grindlay, had challenged the authors to provide examples of specific cases in which plate markings had been useful or necessary in the past. The survey passed this challenge along in the form of questions, "Has anyone done such research with marked plates in the recent past? If so, which plates were important for their work?" For clarity, the responses, itemized below, are divided into two categories—historical and educational uses and scientific uses. This list is representative and should not be construed as all-inclusive. Readers of this report will, no doubt, recall other instances.

**A. Historical and Educational Uses**

The principal historical or educational motives to examine marked plates were to understand or illustrate the work of astronomers and to generate public admiration for the discipline. The plates cited most for such programs were of three varieties: (1) plates on which famous discoveries were marked; (2) plates that illustrated a scientific method; or (3) plates showing a step taken by a particular astronomer in the course of research. Here are some instances:

1. The famous Mount Wilson plate on which Edwin Hubble marked his first discovery of a Cepheid variable in the Andromeda nebula (M31), establishing beyond a doubt that the

M31 was a galaxy outside of our own.  This plate is featured in the astronomy textbook

of Jay Pasachoff and on the Carnegie Observatories website.[17] To wash away a marking

such as this would be tantamount to erasing annotations in copies of Copernicus's *De*

*Revolutionibus*, several people said.[18]

2.  The discovery plates of Miranda and Nereid (the satellites of Uranus and Neptune,

    respectively), taken by Gerard Kuiper in 1948 and 1949 at McDonald Observatory, which

    have these objects marked.[19]

3.  The Harvard College Observatory discovery plate for the Sculptor dwarf galaxy—

    photographed in 1932 with the Bruce 24-inch doublet telescope, then in Bloemfontein,

    South Africa—shows not only the galaxy discovered in 1938, but also observatory

    methods of working with the plate.  Over 2000 galaxies are marked in ink and numbers

    in circles refer to magnitude standards on the plate.  The plate markings are published in

    Gingerich (1990).[20]  (Fig. 5 shows such a heavily annotated galaxy plate.)

4.  A Harvard College Observatory plate showing a portion of the sky near Sagittarius and

    Scorpius with 39 globular clusters circled is published in Harlow Shapley's

    autobiography.[21]

5.  Plates of M101 and M33 that Adriaan van Maanen measured and marked between 1915

    and 1923, showing the direction and magnitude of the rotation of the spiral nebulae,

are illustrated in Robert Smith's history of the debate over the size and nature of the universe.[22] Van Maanen's conclusions on the speed of rotation were accepted by Harlow Shapley as proof that nebulae were within the Milky Way, while they were rejected as erroneous by Heber Curtis because he believed the spiral nebulae were extragalactic and comparable in size to the Milky Way. If they rotated as fast as van Maanen calculated from the plates, the spiral arms would be moving faster than the speed of light. Examination of van Maanen's marked plates has aided astronomers and historians to understand better where his errors originated. (Fig. 6 shows a spiral galaxy's orientation being measured.)

6. A Harvard College Observatory galaxy plate, prints made from plate spectra, and a log book of Annie Jump Cannon, are on public display in *Time, Life, and Matter: Science in Cambridge*, a permanent exhibition of the Harvard Collection of Historical Scientific Instruments. In the exhibition they illustrate the pioneering research methods and scientific life at Harvard College Observatory during the Pickering era.[23]

7. The discovery plate for Pluto taken on 23 January 1930 at the Lowell Observatory is on display in an exhibition, *Exploring the Planets*, at the National Air and Space Museum.[24]

8. Saturn's moon Phoebe, the first moon to be discovered photographically, was found by William H. Pickering in March 1899 by comparing four plates (A3230, A3227, A3228, and A3233) taken on 16-18 August 1898 with the Bruce 24-inch doublet telescope at the Harvard College Observatory station in Arequipa, Peru.  In describing the method used to make the discovery, Edward C. Pickering noted that "in planning the Bruce photographic telescope, a search for distant and faint satellites was regarded as an important part of its work, and accordingly, plates for this purpose were taken at Arequipa".[25]

9.  The Harvard College Observatory discovery plate for Comet Bappu-Bok-Newkirk (C/1949 N1)—J3064, photographed by the 24-33-inch Jewett Schmidt telescope on 2 July 1949— was located by Indian astronomer, Amar Sharma of Nikaya Observatory, Bangalore for his profile of the late Indian astrophysicist M. K. Vainu Bappu in *Biographies of Worldwide Comet Discoverers*.  Images of the marked plate and its jacket have also been published by R. C. Kapoor, and another mistakenly described as the discovery plate but taken on a successive day has been published by Denis Buczynski.[26]

The story of the Comet Bappu-Bok-Newkirk plate is instructive for its historical value and as a demonstration of the usefulness of the markings.   Observing for the first time at Oak Ridge Station of the Harvard College Observatory, Harvard graduate student M. K. Vainu Bappu exposed a photographic plate for 55 minutes with the 24-33-inch Jewett-Schmidt telescope near dawn on 2 July 1949.  Rather than send the plate back to Cambridge for processing, Bart Bok, his professor, suggested that Bappu develop it himself to see what his own plate looked like.  When it came out of the fixing bath, Bappu announced eagerly that he was going to look for comets! "Ha, ha", Bok chuckled.  "Everyone looks for comets".  While the plate was being examined by Bok and Bappu, an undergraduate, Gordon A. Newkirk, Jr. stopped by and was invited to see the good quality of the photograph. Looking at the plate through the binocular microscope, Newkirk exclaimed, "Hey, that looks like the trail of an asteroid or something!" Bok took another look and stated, "That is no asteroid--that is a hairy comet". Two more Jewett plates taken by Bok and Bappu the next night confirmed the discovery.[27]

Harlow Shapley, director of the observatory, announced the new comet,[28] but barely ten days after the discovery, Bappu received a stern letter from the Government of Hyderabad, his sponsor, telling him to stop playing around with comets and get to work on "photoelectric photometry of eclipsing variables". Fred Whipple, chairman of Harvard's Department of Astronomy pushed back, writing in Bappu's defense to the Indian Embassy in Washington, DC, that this was the first occasion in his memory in which a foreign government had seen itself fit to criticize the educational methods of the Astronomy Department of Harvard University. He pointed out that the discovery was accidental to the photographic work that was essential to Bappu's training as a graduate student. "For him to have failed to note this unusual object on his photographic plates would have been a sin of scientific omission; to have failed to announce the discovery would have been a serious neglect of his duty to the scientific world".[29] Whipple continued:

> Our policy of education for graduate students in Astronomy includes thorough background training in classical and positional astronomy, in stellar astronomy, in cosmogony and in modern astrophysics. We will not grant the degree of Doctor of Philosophy to a student who does not have a well-rounded background in all of these areas. If it is actually true that the Hyderabad

Government wishes Mr. Bappu to study "Photoelectric Photometry of Eclipsing Variables" and nothing else in his graduate work, they have certainly erred in sending him to Harvard University. We would be glad to assist him in such a narrow study, if necessary, but we could not grant the degree of Doctor of Philosophy in Astronomy on that basis alone.

Our experience has shown that independence of mind, a broad background in mathematics and the physical sciences and freedom in choosing research problems are essential to a physical scientist who is to produce creative work.

Whipple closed his letter by saying, "Mr. Bappu is doing excellent work as a graduate student….I feel personally that it is a great mistake for him to be handicapped psychologically by ill-founded reprimands that should be directed, if at all, to those who have assumed the responsibility for his graduate education".

So here we have a photographic plate at the center of a great story about a young graduate student who would later be seen as the "father of modern Indian astronomy".  It is a story of youthful exuberance and optimism, serendipitous discovery, instructional methods and educational philosophy, scientific news, and the meddling of foreign powers.  It is a story involving Bok, Shapley, Whipple, Bappu, and Newkirk—all exceptional astronomers.

The whereabouts of this discovery plate were unknown for years to the Indian astronomical community until Amar Sharma was guided by William Liller to the Harvard plate stacks.  Liller, who had been a good friend of Bappu in their student days, observed that if the plate were marked—and the mark saved—it would be a simple matter to find the plate and the comet on it.  Without markings, Sharma would have had to find, look up, or calculate the comet's position, make a finding chart at the appropriate scale, and then match it with the plate images.[30]  As it turned out, there has been confusion in the literature with two different plates published as the discovery plate.  The source of this mistake appears to be the fact that the discovery plate was cleaned and the comet no longer marked (except for a note on the plate jacket), while plates taken on successive nights to confirm the discovery were still marked when a researcher sought them out.[31]

**B.  Scientific Uses**

Astronomers who have used photographic plates in their research offered many scientific reasons to preserve the markings on plates and their jackets.   From their own experiences, they cited:

1. Notations used to identify objects unambiguously in cases where only approximate positions were given in the literature (e.g., plate notations used to re-identify some "lost" variable stars).[32]

2. Notations used to identify the plate referenced in an article only by its date (e.g., "The earliest plate showing the object was taken July 9, 1919...", Solon Bailey describing a particular nova as an "extra-ordinary object" on Harvard photographs "examined by Miss H. S. Leavitt and Miss D. W. Block".)[33]

3. Notations used to identify instrumentation employed when there is no locatable log book and the envelope has no writing, but it is written in the border of the plate image (e.g., early plates at Yerkes with annotations like "M2, 40-inch telescope 1900 Sept 12 8:10 – 11:10". More specifically, a request to identity the telescope used for three Yerkes plates of the Sun, which were lent to Greenwich Observatory in 1939. The information is now needed for a revision of the Greenwich sunspot observations).[34]

4. Notations used to confirm a result or to question an earlier finding (e.g., plates reviewed in order to evaluate the reference frames and stars used by previous investigators, whose parallax measurements disagree with a modern value).[35] The reexamination and measurement of van Maanen's plates of spiral nebulae by modern astronomers is a good example of this.

5. Notations used to find a minor planet and calculate its orbit (e.g., William Liller's and Lola Chaison's work on Minor Planet (2060) Chiron, whose image was found marked in indelible ink on Harvard plates taken in 1943, 1941, and 1897).[36] (See Fig. 7 for a plate illustrating minor planet work.)

6. Notations used to "debug" the New General Catalogue (NGC) and the Index Catalogue (IC) of nebulae (e.g., Harold Corwin's use of HCO plates listed in the *Harvard Annals*, vol. 60).[37]

7. Notations used to identify the plate itself when separated from its envelope.[38]

Respondents suggested that the digital library known as the SAO/NASA Astrophysics Data System (ADS) be searched in order to come up with a starting list of key Harvard College Observatory plates that might merit special attention and preservation with markings intact.

***Can Digital Images Replace the Real Thing?***

The point of this question was to gauge what if anything was lost when the plate was cleaned as part of the DASCH project. Would a digital image of an annotated glass-plate photograph be just as good for the scientific, historical, and educational uses enumerated above as the undisturbed physical plate in its original marked state?

All respondents affirmed that the primary function of the photographic plates, and their reason for being preserved so long, was that they were scientific evidence in time-domain studies. Many understood that a consequence of the ongoing nature of the scientific mission was the erasure of markings—with few exceptions—in order to produce the highest quality digital images of the star fields.[39] But since erasure was irreversible, there were caveats: The photographs of the markings had to be of high fidelity to the originals. The reproduction quality should be checked before any washing was done.  The resolution of the digital images should be no less than 600 ppi, and the images should be shot in "raw" and preserved as tiffs with care taken to color balance the image, something that is not trivial to achieve.[40]  If these conditions were met, this group of respondents thought that photographs of annotations would be an adequate substitute for the original marked plate for most scientific and historical purposes. One respondent with extensive museum and archival experience, David DeVorkin, went further to recommend strongly that 600 ppi prints should be made on acid-free paper and deposited with copies of the annotated envelopes in a suitable archive.[41]  DASCH's Jonathan Grindlay agreed that "this would be nice, and historically appropriate", but pointed out that there has been no NSF funding to replace even a fraction of the acidic envelopes in deplorable condition, much less to make prints on archival paper, and  DASCH has been unsuccessful so far in finding funds from other sources.[42]

A second group of respondents contended that the annotated plates should be scanned, then cleaned, and scanned again, in place of image capture by direct camera photography. "Scanning the markings is a necessity", one wrote, "especially if the image embedded in the emulsion can also be shown in the scan".[43] According to Jonathan Grindlay, this would not be practical.  The annotations on the glass side would not be in focus with the stars on the emulsion side.  Instead, DASCH has promised to enable the superposition of the in-focus, high-resolution photograph of the original plate on the scanned image of the cleaned plate.   This would "be a far more useful historical resource than the original plate, since it would then include all the modern processing (photometry and astrometry) of each and every object".  It would enable the photographic sequence stars used by the original investigators to be checked, systematically, for the first time.[44] Regrettably, this promise remains to be fulfilled.

Although cleaning of the plates seemed unstoppable, and photographs, the only records of annotations that would survive going forward, Jay Pasachoff asked whether it had been shown empirically that the accuracy of the scanning was higher if the plates were cleaned first. "If not, then there is no real gain in the wiping, and the non-wiping method would win overwhelmingly".  He proposed an experiment of taking an uncleaned, unimportant plate and scanning it three times:  first, as is; second after marking it with ink; and third, after cleaning it. Any differences in quality would then be evident.[45]  The authors note that no such experiment has yet been published.

Alistair Kwan, a historian of science then at the University of Rochester, offered another reason to pause and reflect.  Photographs do not reveal as much information as the eyes can in examining an original document.  Paleographers and physically-oriented bibliographers will examine ink thickness, density, texturing of the inky surface, a strain on the substrate, hand writing, fingerprints, and even dirt for clues. These details are only caught by photography if the photographer sets out to capture them with raked lighting, filters, and bracketed exposures.  All this is time consuming.  It is easier for the specialist to examine the original document in order to see its true scale, how it reads at different angles, what properties the medium has, and how reflection and parallax may aid the object handler in better understanding it.[46]

Although most respondents tolerated the erasure of markings, one group remained very unhappy about it, because no photograph would be equivalent.  Some took umbrage at the whole idea of "scrubbing" off any potentially useful annotations.  "Could you find a word other than 'scrubbed'?" one astronomer wrote. "To me this conjures up images of Brillo Pads and Bon Ami ('hasn't scratched yet')....I have visions of Pickering's and Shapley's heroines rolling around in their respective graves sensing that the ink marks left from all those hours and hours of labor might be expunged forever. Like footprints in the sands of time, I suppose".[47]

There were three particular situations in which all respondents agreed that a photograph could not substitute for a marked plate. The first was exhibition. Even respondents who would have washed every plate conceded that those with intact, original annotations would be best for exhibition purposes.[48] The second case was fundraising. "Originals are good for inspiring donors and the general public, and hence for generating funding and other kinds of support. There are memories to be evoked, and great heritage value". [49] The third was calibration and reference. Samples of different types of marked plates should be saved in order to enable researchers to assess the representativeness of the digital images.

### How Many Plates Should Be Saved Uncleaned?

Although respondents disagreed on the number of annotated plates that were worth preserving in their original state, there was consensus that sampling would be adequate. The samples should include:

- Plates illustrating different types of celestial photography, methodology, and annotations on the non-emulsion side (e.g., Fig. 8).

- Plates with different colored inks, along with information about the individuals associated with the inks and their reasons for using the distinct colors. It was also recommended that sample pens, if they survive, should be kept even if they are dried out. Ink differences might be sorted out using chromatography or mass spectrometry.[50]

- Plates that helped establish the variable nature of quasi-stellar objects.[51]

- Plates of the Large and Small Magellanic Clouds with Cepheid variables marked by Henrietta Leavitt.[52]

- Plates marked by Williamina Fleming and other important users, especially the women of Harvard College Observatory.

- Plates used by Harlow Shapley, Adelaide Ames, and others, who studied the large scale distribution of galaxies in the 1920s-1930s.[53]

- Key discovery plates.

Respondents were also concerned about the plate jackets, writing that care should be taken to digitize them as well. If plate rehousing was necessary, then as many as possible of the old jackets should be preserved and archived. Particular attention should be paid to the physical preservation of jackets with markings made by historically significant individuals. At least one or two jackets should be kept for each person involved in the program. [54] If the intent was to throw any digitized jackets away, a number of respondents offered to store them at their home institutions or recommended that they be sold or auctioned off as a fundraiser.[55]

### What Should Be Done with the Scanned Plates?

It was universally agreed by respondents that no matter how exquisite the current imaging equipment is, it can produce "only an observation of an observation".[56] Therefore, the original plates should always be preserved somewhere as a backup, or to look for diffuse veiling (e.g., a very faint supernova light-echo), which the digitizing software may have removed during processing.  Saved plates could also be reimaged in the future as new technology or new ways of interrogating the data becomes available.[57]  The DASCH team completely agreed, and noted that it has been refiling the barcoded plates in their original cabinets after scanning.

The best place to preserve the photographic plates, respondents concurred, was at the institution that created them, since it would presumably have a sense of the historical importance, and it would keep the plates near the log books and working notes associated with them. If housing at the originating observatory or university were not possible, the plates should be stored at a regional or national plate archive. [58]

Wayne Osborn, the author of a census of North American plate collections and a tireless campaigner for their protection, preferred to have "an established plate repository, with several of these scattered around the country and specializing in plates of certain types".  For example, there could be a repository for patrol plates of planetary features, a repository for plates of the Sun and solar eclipses, a repository for slit spectra, a repository for wide-field direct plates, and so forth.  Each repository should have at least one specialist responsible for the plates who would be able to respond knowledgeably to questions about them.  "A problem with repositories", Osborn pointed out, "is that the metadata needed to utilize the plates is often located in notebooks and log books at the observatory or the campus library at which the plates were taken.  Sending the plates to a distant repository likely means separating them from the log books".  This problem could be solved, however, by having images of the log book pages available online, readily searchable, and linked somehow to the plate images.[59]

Who would fund the programs of plate preservation and storage?  Since the plates exist as astronomical data, the responsibility for their care and maintenance resides with the astronomical community, not with historians.[60]

Most respondents agreed, however, that some photographic plates deemed to have high historical value could be loaned or transferred to museums for display purposes or historical preservation in their collections.  Nonetheless, Elizabeth Griffin, an astrophysicist at the Dominion Astrophysical Observatory and Chair of the IAU Task Force for the Preservation and Digitization of Photographic Plates, observed that she "would not recommend separating parts of any high-quality collection for museum status unless the owners of the digital records are happy to allow it".[61]

**Philosophical Points Raised**

Respondents to the survey were very thoughtful and raised some interesting philosophical issues.  Even those with strong opinions recognized that "there may be no clear-cut answers", but only disparate visions of preferred practices, the outcome of which "must perforce be conditioned by limited resources".[62]

Elizabeth Griffin spoke for many astronomers in voicing concerns about drawing a line between old things and historical things. "What is it that transforms an artefact into something of heritage value? Is it historic just because it is old?" Griffin asked. She noted that many of the people who had made the telescopic observations and plate markings were just doing their jobs, and that there was nothing magical about the way they did it. Could it be that the passage of years had mysteriously hallowed their work? "We would not so readily attribute the same sanctity to data observed last night by our contemporaries". Griffin suspected that the decisions were "more emotional than objectively scientific", and that rarity played a part.[63] To this, a historian would reply that the decisions were not based simply on age or rarity or emotion, but on ways in which the markings showed us how business was done then. Moreover, the boundary between historic and non-historic is not a rigid one drawn between dichotomies, nor does scientific objectivity (whatever that may be then or now) play a part in making the distinction. The essence that makes an object historical is fluid and varies in the context in which it is seen and interpreted.[64]

In selecting plates to preserve uncleaned, how do we know when "hand-written marks or annotations are seriously scientific or only somewhat casual?"[65] Historians would concede that it is difficult to ascribe motivations to past actors on the basis of the tangible remains of their actions alone. Nonetheless, the difficulty in figuring this out does not make it impossible to do so; there may be other documents and objects that shed light on the motivation for and importance of particular markings.  Moreover, the historical value of any particular item may have little to do with whether it was created in a casual or calculated manner.

Indeed the routine nature of an activity may be good reason to preserve evidence of it. One astronomer reminisced, "It was not uncommon to write on plates.  I did so when I observed spectra photographically.  Quite often we wrote in pencil on the emulsion *before* exposing, for purposes of identification, so as to render it indelible after developing.  We might also put ink blobs to indicate (to ourselves or an assistant) the wavelengths at which the intensity calibrations should be traced, or arrows to remind ourselves (or an assistant) which was the spectral region of particular interest".  Her conclusion was, "Markings like that have no historic value, and should be erased".[66]  In contrast, a historian of astronomy finds such information priceless in understanding scientific practices, even if the work is mundane to its practitioners.  What we have here is tacit knowledge revealed.  The recent book, *Observing by Hand* by Omar Nasim, makes this point forcefully.   The unpublished sketches and annotations of nebulae found in astronomers' personal observing notebooks reveal how the observers shaped and constituted celestial phenomena and the processes they employed in the production of scientific knowledge.[67] Certainly not every such marked plate need be protected, nor should they all be dismissed out of hand.

"When all is said and done, we must also not lose sight between the two opposing properties: the objective and subjective", Griffin wrote. "A photographic plate is an objective observation, but the marks which were added afterwards may only be someone's opinion, guide or aid, and are definitely subjective. The scientific process depends on retaining the objective and keeping a clear boundary between it and the subjective".[68]  Here Griffin called our attention to a much debated conundrum at the heart of investigations into the natural world for thousands of years—i.e., the boundaries between objectivity and subjectivity in empirical science.  And she showed us that the humble astronomical photographic plate sits at that boundary.  Philosophers and historians of science would say that subjectivity enters the story before the plate is exposed, developed, and read.  The decision to photograph that part of the sky with that instrument and wavelength sensitivity is already colored by human thought and culture.[69]  Taken to the logical conclusion, there are no unvarnished facts.  But of course, scientists, as a matter of practice and practicality, believe that they work in a world where their observations can and should be objective and not influenced by their beliefs.

**Concluding Thoughts**

The results of the survey show that many eminent astronomers and historians firmly believe that annotations on the photographic plates and their jackets have significant scientific, historical, and educational value.  As authors, we agree.  We endorse the preservation of as many markings as possible.  In cases where current scientific needs can only be met with the permanent removal of annotations, we urge scientists to have technicians photograph or scan the plates and jackets using high resolution, color-balanced protocols and archiving procedures.  To assist observatories and plate repositories facing this daunting task, we append to this paper the procedural recommendations of conservators, photographers, and archivists on plate handling, cleaning, photography, and rejacketing.

We would also like to point out that the arguments raised in this paper are pertinent in other disciplines and areas within the academy.   This is not just a story about astronomy and its photographic data.

First, there is a growing body of work by historians on the history of note-taking and annotations.  This scholarship examines notes not just for clues of the development of a particular writer's thoughts, but also to understand collective practices that have defined professions, occupations, audiences, cultural groups, times and places, and purposes.  The content and material culture of notes can show us how ideas and data are shaped and transmitted, organized and preserved, solidified, disputed, and repudiated.[70] Closely associated with note-taking are drawings, sketches, maps, arrows, circles, and assorted visual languages.  Scholars have begun to examine these markings too, and see in them further evidence of annotations focusing one's attention and memory, and being intimately entangled with observation, interpretation, and objectivity.[71]  In the words of Lorraine Daston, by looking, writing, and drawing, nature is "made intelligible by being made legible".[72]

Second, there is a substantial body of literature that discusses the history and problematic nature of scientific photography, showing that it has never been as isolated and objective as its early proponents claimed.   In the nineteenth century, scientists welcomed the camera for its "mechanical objectivity", even as they worried about the instability of photographic methods, chemical emulsions, and their sensitivity to light and color.  It mattered how photographic plates were prepared, developed, and stored.  For instance, as collodion-process wet plates dried, their emulsions could warp and buckle, producing distortions in the image, making them unreliable for precise measurements.  The choice of photographic lenses, apertures, and focal lengths also affected the resultant image's flatness, scale, and exposure time.  Astronomers debated the pros and cons of daguerreotypes, wet plates, and the new gelatin dry plates in advance of their national expeditions to observe the Transit of Venus in 1874. In practice, scientific photography by different individuals of the same subject—whether a Transit of Venus, a solar eclipse, or a horse race—could be as different as written reports or drawings of observations, leading historians and philosophers to ask if it might be better to talk about 'making' rather than 'taking' a photograph. Indeed, recent scholarship has shown that skill, taste, social class, gender, politics, and belief undermine the objective and evidentiary claims of photographs.[73]

If it's a fallacy to claim that photographs are witnesses of reality unmediated by human agency, then why should we treat them as different from written sources? We can ask the same questions about how they were made, to whom they circulated, and what meanings they had to those who examined them. We must conclude, therefore, that astronomical photographs should not be privileged over their annotations, jacket manuscripts, and associated logbooks. All must be studied together, if historians of science are to have a complete picture of the way astronomers actually used photography in their work.

Lastly, what seems to be at stake here is the nature of the archive and the power of its diverse constituents. The erasure of plate markings would certainly be a loss to historians but also to scientists, as the results of this survey show. The driving force for such erasure today is a narrowly focused interpretation of the needs of long-time-domain astronomers. But why should this group be permitted to act on a judgment that will irreversibly alter the state of the archives for everyone else? Who gets to decide? Indeed, even if altering the artifacts seems unproblematic now, should we let our lack of imagination foreclose prospects of their use in the future? We may not yet have "invented the technologies or articulated the conceptual frameworks that might reveal what is latent in them today", one respondent remarked.[74] "We cannot say today how science and history might develop tomorrow".

(CHSI), Harvard University included Dr Jean-François Gauvin, Director of Administration; Dr Sara Schechner, David P. Wheatland Curator; Martha Richardson, Collection Manager; and Samantha van Gerbig, CHSI photographer and senior curatorial technician.  Consultants from the Harvard Library were Dr Frenziska Frey, Malloy-Rabinowitz Preservation Librarian, and Head of Preservation and Digital Imaging Services; William Comstock, Head of Imaging Services; Andrea Goethals, Manager of Digital Preservation and Repository Services; Brenda Bernier, James Needham Chief Conservator and Head of the Weissman Preservation Center; Elena Bulat, Photograph Conservator for Special Collections; Anabelle Chabauty, intern at the Weissman Preservation Center; and Jacqueline Ford, imaging assistant at the Wolbach Library. Others included Dr Owen Gingerich, Professor Emeritus of Astronomy and the History of Science, Harvard University; Dr Elizabeth Griffin, Dominion Astrophysical Observatory, Chair of the IAU Task Force for the Preservation and Digitization of Photographic Plates; and Dr Jay Pasachoff, Field Memorial Professor of Astronomy and Director of the Hopkins Observatory at Williams College, Chair of the Historical Astronomy Division of the American Astronomical Society.

[3]In correspondence with the authors, 29 December 2014, Jonathan Grindlay claimed that "testing of the wire-brush cleaning method required to remove ~100y old India ink revealed it is physically impossible for a stainless steel wire brush to scratch glass, for the obvious reason that the hardness index of glass exceeds (significantly) that of steel!"  He plans to publish these results.   Bart Fried, Principal, All Glass International LLC, Forest Hills, NY pointed out, however, that surface hardness and scratch resistance are not the same thing, that there are diverse tests

to measure each of these, and that Grindlay's claim does not take into account any particulates in the grime that will be pushed around by steel brushes.  Bart Fried to the authors, 20 July 2015.

[4] The 2013 reports reiterated advice given 6-12 months earlier by the professional staff of the Collection of Historical Scientific Instruments and the Weissman Preservation Center at Harvard to DASCH:  (1) Samantha van Gerbig, "Comments on the HCO DASCH plate scanning project", 31 May 2012; a 10-page report of best practices in museums and archives for plate handling, cleaning, lighting, color balancing, image capture, and metadata, submitted by the CHSI photographer and endorsed by Jean-François Gauvin and Sara Schechner, respectively the CHSI administrative director and curator.  (2) David Sliski, "Advice from a meeting with Brenda Bernier, Paul M. & Harriet L. Weissman Senior Photograph Conservator, and Elena Bulat, Photograph Conservator of the Weissman Preservation Center, Harvard Library, January 25th 2013", DASCH project notes, January 2013.

[5] For example, see W. Osborn and L. Robbins, *Census of astronomical photographic plates in North America: Final report* (Washington, D.C.: American Astronomical Society Working Group on the Preservation of Astronomical Heritage, 2008), https://aas.org/files/censusreport-final.pdf.  A *Workshop on Developing a Plan for Preserving Astronomy's Archival Records* was held on 18-19 April 2012 at the offices of the American Institute of Physics, College Park, MD, USA.  It was organized by the AAS Working Group on the Preservation of Astronomical Heritage (WGPAH) and co-sponsored by the American Astronomical Society and the American Institute

of Physics with support from the National Science Foundation.  The report of the meeting, *Preserving astronomy's North American heritage records* is published online by WGPAH:

https://aas.org/files/wgpah_april_2012_workshop_report.pdf.

[6] David DeVorkin, email to the AAS Working Group on the Preservation of Astronomical Heritage (WGPAH), 9 June 2013.  DeVorkin expressed his own opinion and not that of the Smithsonian Institution.

[7] Brad Schaefer to HASTRO-L, 7 June 2013; and Brad Schaefer to David Sliski, 18 June 2013.

[8] See Osborn and Robbins, *Census of astronomical photographic plates* (ref. 5); and Wayne Osborn *et al.*, "Making archival data available for research in the next decade and beyond", pp. 160-165, in *Preserving astronomy's photographic legacy: Current state and the future of North American astronomical plates*, ed. Wayne Osborn and Lee Robbins, ASP Conference Series, ccccx (San Francisco: Astronomical Society of the Pacific, 2009)  (ADS Bibcode: 2009ASPC..410..160O).

[9] Email exchange between Jay Pasachoff, Wendy Freedman, Barry Madore, and Sara Schechner, 28 May 2013.

[10] William Liller to Sara Schechner and Jonathan Grindlay, 19 June 2013.

---

[11] Sponsored by the National Science Foundation, the PARI workshop on 1-3 November 2007 brought together individuals responsible for observatory plate collections to discuss ways to prevent old plates from being discarded or stored under poor conditions.  For recommendations stemming from the workshop, see Wayne Osborn and Lee Robbins, "The workshop on a national plan for preserving astronomical photographic data", and M. W. Castelaz, "The Astronomical Photographic Data Archive at the Pisgah Astronomical Research Institute", pp. 33-69 and 70-78 in *Preserving astronomy's photographic legacy: current state and the future of North American astronomical plates*, ed. Wayne Osborn and Lee Robbins, ASP Conference Series, vol. ccccx (San Francisco: Astronomical Society of the Pacific, 2009) (ADS Bibcodes: 2009ASPC..410...33O and 2009ASPC..410...70C).

[12] Vladimir Strelnitski to David Sliski, 12 June 2013.

[13] Peter Boyce to Sara Schechner, Joe Tenn, and Jonathan Grindlay, 9 June 2013.

[14] Virginia Trimble to WGPAH, 6 June 2013; Lee Robbins to Sara Schechner, 10 June 2013; Rocky Kolb, "Be  careful what you rub out", 504th Convocation Address, University of Chicago, 27 August 2010,

https://convocation.uchicago.edu/sites/convocation.uchicago.edu/files/uploads/504th%20-%20Rocky%20Kolb.pdf; Allan Sandage, *Centennial history of the Carnegie Institution of Washington*, vol. i, *The Mount Wilson Observatory* (Cambridge: Cambridge University Press, 2004), 495-498.  According to Robert W. Smith, *The Expanding Universe: Astronomy's "Great*

Gallery, 26 August 2013, http://britastro.org/gallery_image/2595. Buczynski published plate

J3068, taken 3 July 1949, which is mistakenly described as the discovery plate. J3068 has the comet and reference stars marked clearly in red ink, whereas the discovery plate J3064 no longer has the comet marked, making it very hard to find. This illustrates the confusion that may arise when annotations have been cleaned off.

[27] "Harvard Tyro Finds Comet", *The Christian Science Monitor*, 6 July 1949,

http://search.proquest.com.ezp-prod1.hul.harvard.edu/docview/508082730?accountid=11311

(accessed 8 December 2014); Kapoor, "Comet Bappu-Bok-Newkirk" (ref. 26).

[28] Harlow Shapley, "New Comet" and "Comet Bappu-Bok-Newkirk", Harvard College

Observatory Announcement Cards 1006-1008, 5, 8, and 11 July 1949, John G. Wolbach Library,

Harvard-Smithsonian Center for Astrophysics.

[29] Fred L. Whipple, letter, 26 July 1949, quoted in full in Kapoor "Comet Bappu-Bok-Newkirk"

(ref. 26), 118.

[30] William Liller in email to Sara Schechner and Jonathan Grindlay, 19 June 2013; A. Sharma,

"Tracing the photographic plate of Comet Bappu-Bok-Newkirk", *Current science (Bangalore)*, cv,

no. 3 (10 August 2013), 295-6, *Academic Search Premier*, EBSCO*host* (accessed 9 December

2014). Grindlay hopes that DASCH's high resolution, color-corrected photographic image of each plate and its barcoding system will make the process of identification and location in the Plate Stacks easier than it has been previously.

[31] *Cf.* Kapoor, "Comet Bappu-Bok-Newkirk" (ref. 26) and Buczynski, "C/1949 N1 Bappu-Bok-Newkirk" (ref. 26).

[32] W. Osborn and O. F. Mills, "The Ross variable stars revisited. I." *Journal of the American Association of Variable Star Observers*, xxxix, no. 2 (2011), 186 (ADS Bibcode: 2011JAVSO..39..186O) and "The Ross variable stars revisited. II." *Journal of the American Association of Variable Star Observers*, xl, no. 2 (2012), 929 (ADS Bibcode: 2012JAVSO..40..929O). Another example of the use of plate and envelope notations was a case where a particular star was found to have high proper motion, but the comparison stars were in a variable star study. In a note to the authors, 29 December 2014, Grindlay responded, "Now the precise RA, Dec positions of each object resolved on each plate are produced by the DASCH scans – complete with proper motion corrections applied. And the photographic images of the original plate make it just as clear how these objects relate to the historical record, but now with far better positions and corrections to compare with other historic or current/modern studies".

[33] S. I. Bailey, *Harvard College Observatory bulletin*, no. 680, 1919 (ADS bibcode 1919BHarO.680....1B); Wayne Osborn did find the key Harvard plate referenced in the article. See G. Luberda and W. Osborn, "New light curve for the 1909 outburst of RT Serpentis", *Journal*

*of the American Association of Variable Star Observers*, xl, no. 2 (2012), 887. (ADS bibcode: 012JAVSO..40..887L)

[34] Wayne Osborn to WGPAH, 6 June 2013: "We identified the telescope by locating three plates taken on the known dates for which the notation 'returned from Greenwich Observatory 1939' was on the plate jacket." Wayne Osborne to Sara Schechner, 19 December 2014.

[35] Jennifer Bartlett to WGPAH, 23 August 2014; Peter Broughton to HASTRO-L, 7 June 2013: "There are surely situations where an astronomer has unwittingly misidentified a star on a plate and subsequent researchers would find this misidentification helpful in re-examining the astronomer's analysis."

[36] William Liller to Sara Schechner and Jonathan Grindlay, 19 June 2013: "Discovered by C. T. Kowal in 1977, its [Chiron's] orbit, later calculated by B. G. Marsden to be mainly between those of Saturn and Uranus, was only crudely known until yours truly with Lola Chaison found its image on Harvard plates taken in 1943, 1941 and 1897. The 1941 image was found on one of those absolutely magnificent 14" x 17" Bruce astrograph 3-hour exposures; it was located roughly a half degree from its guesstimated position. The image, short and faint, was, in fact, marked (in indelible ink) because the plate inspector thought it was that of a faint distant galaxy seen edge-on. (As I remember, there were more than a thousand images marked on that plate.) One can imagine the difficulty in relocating that image if all the marks were scrubbed off." See C. T. Kowal, W. Liller, and B. G. Marsden, "The discovery and orbit of /2060/ Chiron",

pp. 245-250 in *Dynamics of the Solar System*, ed. R. L. Duncombe, IAU Symposium Series No. 81 (Dordrecht: D. Reidel Publishing Co., 1979) (ADS Bibcode: 1979IAUS...81..245K).

[37] A list of HCO plate numbers on which IC objects were found is in the *Harvard Annals*, lx, 152-153, 178.  A few additional plates with asteroid trails noted are on pp. 176-177.  Harold Corwin wrote that this debugging has been a "hobby" of his for over forty years:  "Many IC objects were discovered on Harvard plates taken in the 1890s and early 1900s….A few of the IC objects have not been found on subsequent plates:  Are these missing objects plate defects, or possibly unresolved multiple star images?  Perhaps their positions were simply recorded incorrectly.  The [annotated] plates may help provide answers."  Harold Corwin to Sara Schechner, 22 August 2013.

[38] Wayne Osborne to Sara Schechner, 19 December 2014, citing his experience at Yerkes.

[39] Elizabeth Griffin to WGPAH, 7 June 2013; Jennifer Bartlett to WGPAH, 23 August 2013; and Tom Corbin to Sara Schechner, 12 June 2013.

[40] Van Gerbig, "Comments on the HCO DASCH plate scanning project" (ref. 4);  Sliski, "Advice from a meeting" (ref. 4); David DeVorkin to WGPAH, 9 June 2013; Elizabeth Griffin to WGPAH, 7 June 2013; Tom Corbin to Sara Schechner, 12 June 2013.

[41] David DeVorkin to WGPAH, 9 June 2013.

[69] See Jessica Ratcliff, *The Transit of Venus enterprise in Victorian Britain* (London: Pickering & Chatto, 2008); Jessica Ratcliff, "Models, metaphors and the Transit of Venus in Victorian Britain", *Cahiers François Viète*, xi-xii (2007), 63-82; and Alex Soojung-Kim Pang, *Empire and the Sun: Victorian solar eclipse expeditions* (Stanford: Stanford University Press, 2002), chap. 4: "Drawing and photographing the corona".

[70] Ann Blair, "Note taking as an art of transmission", *Critical inquiry*, xxxi, no. 1 (2004), 85-107; Ann Blair, "Humanist methods in natural philosophy: The commonplace book", *Journal of the history of ideas,* liii (1992), 541-551;  Ann Blair and Richard Yeo, eds., *Note-taking in early modern Europe*, special issue of *Intellectual history review*, xx, no. 3 (2010);  Ann Blair, "Student manuscripts and the textbook", pp. 39-73 in *Scholarly knowledge: Textbooks in early modern Europe*, eds.  Emidio Campi, Simone de Angelis, Anja-Silvia Goeing, and Anthony Grafton (Geneva: Droz, 2008); Lorraine Daston, "Taking Note(s)", *Isis,* xcv (2004), 443-448;  Michael R. Canfield, ed., *Field notes on science and nature* (Cambridge: Harvard University Press, 2011); Frederic L. Holmes, Jürgen Renn, and Hans-Jörg Rheinberger, eds., *Reworking the bench: Research notebooks in the history of science* (Dordrecht: Kluwer, 2003); Richard Yeo, "Notebooks as memory aids: Precepts and practices in early modern England", *Memory studies*, i (2008), 115–136. See also the conference at the Radcliffe Institute for Advanced Study, *Take note* (2012), http://www.radcliffe.harvard.edu/event/2012-take-note-conference.

# Preservation Recommendations for Historic Glass Astronomical Plates

**Brenda Bernier**
Weissman Preservation Center, Harvard Library, Harvard University, USA

**Elena Bulat**
Weissman Preservation Center, Harvard Library, Harvard University, USA

Glass astronomical plates are very similar to glass plate negatives in material composition, deterioration, and preservation. Most plates are comprised of soda-lime-silica glass coated with a gelatin emulsion in which silver image particles have been developed. Plates were commonly annotated on the glass side with black or colored inks. Based on the physical and chemical properties of these historic objects, the following is recommended:

## Storage and Handling

- Use unpowdered nitrile gloves and handle the plates by the edges. Do not use cotton gloves because the plates will be too slippery to handle safely. Fingerprints left by ungloved hands can permanently etch into the emulsion or the glass.

- Broken plates should be protected on the emulsion side with a piece of borosilicate glass cut to the same size. A second piece of glass can be placed directly next to the broken glass, but this is rarely needed. Borosilicate glass is chemically more stable than soda-lime-silica glass. It is very difficult to cut so should be purchased pre-cut to standard or custom sizes. Secure the support glass to the plate by binding all edges with an appropriate, archival, self-adhering tape. The chosen tape should be easy to work with and have passed the Photographic Activity Test (PAT), which ensures that it will not cause fading or staining of the silver image over time. Commercial tapes can change formulas without notification from the manufacturer. For ease of handling when binding broken plates, clean bare hands can be used with great care, in lieu of gloves.

- Poor quality enclosures can cause fading or staining of the photographic image. Differential fading is highly possible and would be nearly impossible to detect on astronomical plates, leading to inaccurate analysis. The latest international standard for photograph enclosures[1] recommends buffered enclosures that pass the Photographic Activity Test; archival suppliers will indicate which products pass the PAT. This standard applies to both the envelopes used for each plate as well as to storage boxes, if used.

- Stacked plates will crack under their own weight. Store plates vertically.

- Environmental pollutants can fade or stain the image. As with poor quality enclosures, differential fading is possible. Warm environments, particularly with high or fluctuating humidity, can speed the deterioration process or even cause the emulsion to separate from the glass. The latest international standard for storage of glass plate negatives[2] recommends a maximum temperature of 18°C and a relative humidity range of 30-40%.

## Preparation of Plates for Scanning

- NEVER clean plates that have flaking emulsion; handle these emulsion side up only.

- For plates in good condition, the first type of dusting to try is using an air bulb dust blower. This is gentle enough to use on either the glass or emulsion side. Do not use canned air because it contains accelerants which can damage the emulsion. If the air bulb is not sufficient, try dusting with a soft brush on the glass or emulsion side. Wash the brush nightly. Use extreme caution when dusting emulsion side as grit can cause scratches.

- For plates in good condition, the GLASS SIDE can be wiped with a soft, lint-free cloth or disposable wipe. Commercial glass cleaners are to be avoided as they contain ammonia and other chemicals that can fade the silver image. If some moisture is needed, try blowing on it like when cleaning eyeglasses then rub with soft cloth ("huff and buff").

- If more cleaning is necessary, try a water/ethanol solution on the GLASS SIDE only. Do not use moisture or solutions on emulsion side. Gelatin emulsion is very sensitive to water and can easily swell with moisture; therefore try to use the least proportion of water in the solution as workable (probably 40:60 water/ethanol) and use the smallest amount of liquid for cleaning. Ideally, the solution would be dropped or sprayed onto the lint-free cloth and not applied directly to the glass.

- As part of the scanning workflow, the institution may desire the removal of the historic annotations. If that is so, high quality photographic documentation should be taken before cleaning and removal of the annotations. In fact, the American Institute for Conservation standards for practice specifically states that photographic documentation is the minimal standard of practice for "those aspects that may be altered by the treatment."[3]

- If ink annotations must be removed, the conservation recommendation is to clean the glass with a water/ethanol solution as described above. Razor blades can be carefully used, but do pose a risk for scratching the glass. Wire brushes are not recommended because the risk of scratching the plate is high.

## Notes on Contributors

Brenda Bernier is the James Needham Chief Conservator and Head of Weissman Preservation Center, Harvard Library, where she oversees the preservation of rare books, works on paper, and photographs. Brenda holds a M.S. in photograph conservation from the University of Delaware. She has written and presented extensively on the care of photographic collections.

Elena Bulat is a photograph conservator at the Weissman Preservation Center, Harvard Library. She is a graduate of the Advanced Residency Program in Photograph Conservation at the George Eastman Museum. Her current research is on refining analytical techniques to identify coatings on photographs in order to inform preservation strategies and to deepen art historical understanding of early photography.

## Notes

1. ISO 18902:2013

2. ISO 18918:2000

3. http://www.conservation-us.org/about-us/core-documents/commentaries-to-the-guidelines-for-practice/25#.VKyHpivF-EU

# Photography Recommendations for Historical Glass Plate Negatives

**Samantha van Gerbig**
Harvard University, USA

**David Sliski**
University of Pennsylvania, USA

One of the cardinal rules of museum and archive management is that objects should only be altered or modified using methods that are reversible. Occasionally, this rule may have to be disregarded. When these rare circumstances occur, collection managers are obligated to ensure that any information that will be altered or removed is preserved beforehand in the best possible manner available at the time.

In regards to glass photographic plates that contain potentially significant astronomical data, there are times when historic annotations (usually made in India ink) may need to be erased before that data can be accurately extracted from the negative. This report offers advice on how to digitally capture these markings before they are permanently destroyed. Methods for capturing handwritten documentation from the paper jackets that protect glass plates are also addressed.

## Workstations

Documenting the annotations on glass plate negatives and their jackets requires ergonomically-designed workstations, with two cameras and different lighting setups: one for capturing the ink markings on the plates via transmitted light, and the other for shooting the plate jackets with standard illumination.

Camera height should always be adjusted to ensure that the image of the plate to be photographed fills the camera frame. This practice maximizes the amount of detail captured by the camera for each plate.

When setting up the copy stands for the first time (and every time camera height is altered thereafter), it is crucial to ensure that the plane of both the camera sensor and light box are parallel. This can be checked via a bubble level.

### Plate Jacket Station

A copy-stand setup with two lights (one on either side) should be used for this station. It can be custom-built, off-the-shelf, or refurbished equipment. Bencher of Antioch, IL makes high quality units with interchangeable lighting fixtures. Refurbished darkroom enlarger stands fitted with aftermarket lighting setups are sometimes a more affordable alternative. LED studio lights are becoming more commonly used; one maker is IKAN.

## Plate Marking Station

To reduce operator eye fatigue the light box used at this station should be high quality. At the time of this writing, some of the best on the market are made by the German company JUST Normlicht, who produce a range of light boxes that use electronic dimming control and ballast to eliminate flickering. LED light boxes with ground glass tops are becoming more popular and allow long-term color temperature stability.

Black window or L-shaped mats can be made in different sizes to fit around plates as they are photographed. These can be inexpensively made at any picture framing shop out of acid-free cardstock, but these wear out often. A custom black plastic guide is more durable.

The surface of the light box must be kept clean, as dust from degrading jackets and flaking emulsion can scratch the emulsion surfaces of the plates. Short blasts of clean, dry, compressed air (not "air-in-a-can") are recommended for this. This station should be cleaned daily to ensure safe working conditions for all plates for the duration of any project.

A sheet of 3-mil archival-quality polyester film should be placed between the emulsion side of the plate and the surface of the light table during shooting. This will protect both surfaces from scratches. These sheets should be discarded when they are scratched or otherwise dirtied. This material comes in many different sizes and is available from museum conservation supply firms.

## Computer Workstation

A powerful computer with more than 16 GB of RAM and a 2 GB graphics card is crucial for digital image processing and management. A DVD or Blu-ray drive will be useful for burning read-only copies of the files for long-term storage. Common software used includes Adobe Bridge, Photoshop, and Lightroom.

Monitor quality is important. Since computer monitor color temperatures vary widely (for example, Windows monitors range between 7300-9300K), the workstation monitor must be calibrated. This process works by setting the monitor to a neutral output based on the lighting in the workspace. It also measures the monitor's native imperfections and produces a filter to compensate for those imperfections. Since these values can vary over time, many professional photographers recalibrate their monitors monthly. A wide range of tools for this process exist. As of this writing, a well-regarded system is the X-Rite i1 Display Pro.

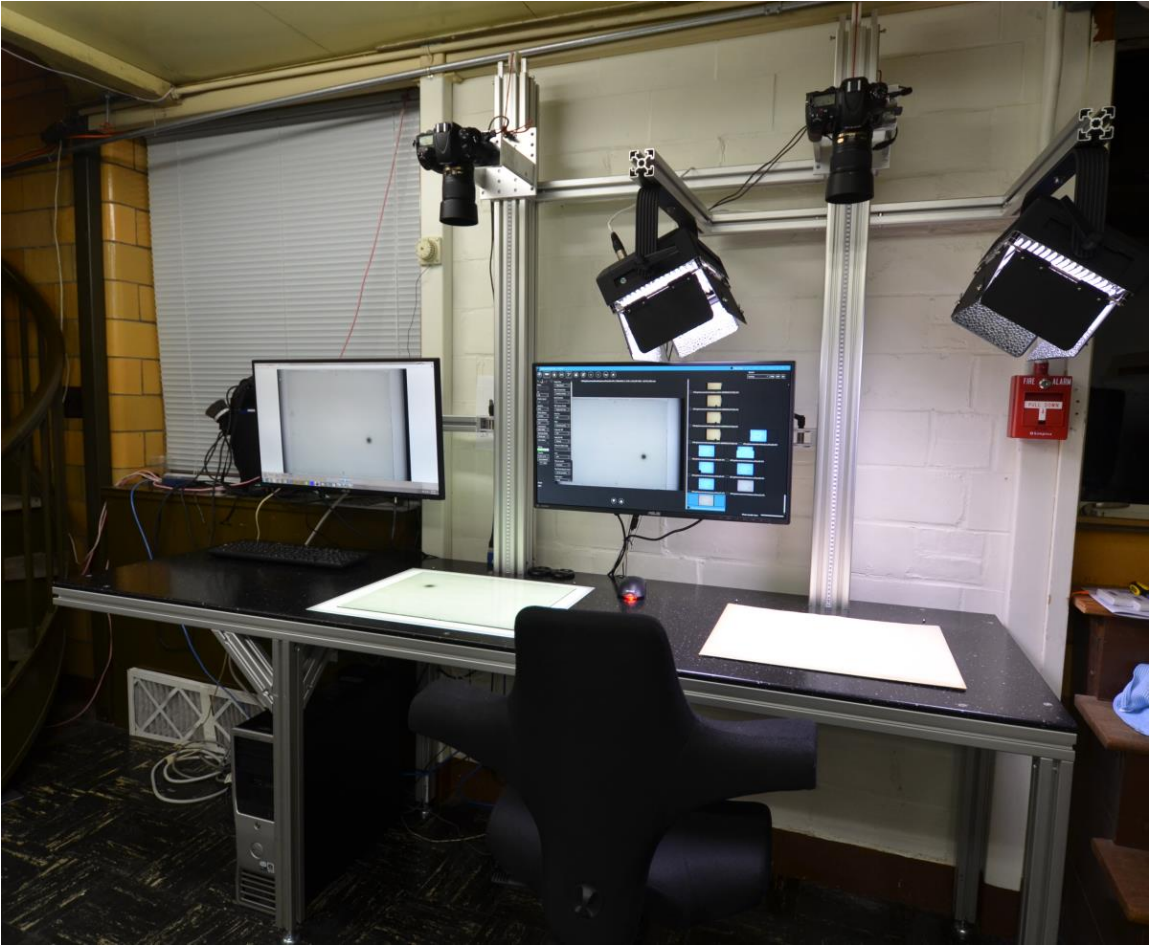

**Figure 1.** At Harvard College Observatory, the DASCH photography station features two Nikon DSLRs with 60mm macro lenses on height adjustable pillars. The LED lights used for reflective photography are also height and width adjustable. To move a camera vertically, the operator loosens a hand screw. When the tension is released, the camera's position is maintained by a counterweight located in the pillar. This prevents an operator from dropping the camera and allows for easier use. A measuring tape permits the camera to be reset to a specific location. Both cameras are positioned over the center of the light source. *On the left*, the transmission photography station features a 16" x 20" lightbox that uses LEDs with a ground glass top, which evenly disperses the light. *On the right,* the reflective media station features two IKON LED lights, which are used to illuminate the jackets. (Not shown are polarizing filters over both lights and a rotating polarizing filter on the camera lens.) Both the plate and jacket stations have accompanying laser-cut plastic guides, which allow for easy positioning of each jacket and plate for high-throughput photography. The monitor on the left shows each image captured by the plate camera; the central monitor shows each jacket image. A pair of solid state relays (not visible) turn off the monitors during capture. Additionally, as one camera is taking a photograph the other camera's light source is off to ensure a singular, even color balance for each picture.

## Cameras and Lenses

Each copy stand should be equipped with the highest-quality cameras that the project can afford. For this type of project, good-quality results can be had using consumer to professional-level SLR cameras. Minimum camera resolution should be no lower than 25 megapixels. At this time, the highest resolution consumer cameras available are Nikon's D810 (36.3 MP) and Canon's EOS 5Ds (50.6 MP). For plates larger than 8x10, a medium format camera is recommended.

For documenting plate markings, the highest resolution achievable within the budget of the project is recommended (600 PPI is a good value to aim for).

Whatever the camera used, it should be equipped with a fixed focus lens with a focal length longer than 50mm. This type of lens displays less barrel distortion than other lenses, particularly zoom lenses, which should absolutely be avoided for copy stand work. If the copy stand is tall enough and the room height allows it, the use of a short macro lens such the Nikon 60mm f/2.8 will reduce edge distortion to a minimum.

If a macro lens is available or affordable, it can be used for close-up shots of faint or otherwise hard-to-distinguish historical markings. If using a Nikon camera body, the AF-S 105mm f/2.8 VR Micro-NIKKOR lens is recommended.

To minimize digital noise from the camera's sensor, the ISO setting should be kept as low as possible. A range of 64 to 140 is ideal.

Exposures should be adjusted via changes in shutter speed. Bracketing tests should be used to make sure that the correct exposures are used for different media. If your lighting system produces enough wattage, f-stops should be kept in the f/8 range.

A tethered capture system, wireless trigger, or a cable release should be used to minimize any chance of images blurring from the motion of the photographer. Shutter speeds should be shorter than 1/30 of a second.

Camera lenses should be outfitted with UV filters. These are inexpensive and protect front lens elements from dust and scratches.

If in constant use, cameras should be cleaned and serviced at least once a year by a shop authorized by your camera manufacturer.

## Lighting, Color, and White Balance

### Copy stand lighting

For even illumination, copy stand lights should be aimed so that the right-hand fixture illuminates the left-hand side of the jacket to be photographed and vice versa. Strive to eliminate glare. This can be done by keeping the lights lower than 45° from the vertical and by using polarizing filters on the lights in concert with an adjustable polarizing filter on the camera lens. If a polarizing filter is used on a camera, it should replace the UV filter.

## Color Balance

It is important that the plates and jackets documented via this archival process are photographed in a manner that strives for accuracy, both in detail and color. With this in mind, it is ideal for the work to take place in a darkened room or a room where the lighting can be controlled from a single source. Daylight is to be avoided at all costs as it will make color balance impossible.

Within this space, a single *type* of light source (i.e., LED, incandescent, fluorescent, etc.) should be used to illuminate the plates at both workstations. Doing this will insure consistency in *color temperature* across the span of the project. This value, measured in degrees Kelvin, refers to the hue of a light source, which is directly related to its wavelength. Different types of lighting can have a wide range of color temperatures. Here are a few average values:

- Incandescent            2,900°K
- Fluorescent              5,000°K
- Daylight (overcast day)     6,500°K
- Daylight (sunny day)       8,000°K

Images photographed in longer-wavelength, low-temperature light will have a "warm" yellow-orange cast, while those taken in shorter-wavelength, high-temperature light will have a "cold" bluish cast.

The goal with color balancing is to ensure that any hue imposed by lighting can be accommodated by the camera (it is very difficult, if not impossible, to compensate in the camera for mixed color temperature lighting). This calibration process only works well if the same type of lights or lighting with the same color temperature is used across the entire workspace. In the end, incandescent, fluorescent, and LED lamps can all be used, just as long as all sources emit light at the same color temperature.

Most lamps come from the manufacturer with reliable listed color temperatures. To check whether these values are correct, a camera can be used to measure the color temperature of any light source. This can be done as part of the white balancing process (see below).

If possible, the monitors should be turned off during the capture of the image as that adds a light source that varies during shooting. This can be done with simple electronics and solid state relays such that when the camera trigger is activated the monitors will switch off.

Color targets should be included in all reflected-light photography, as recommended by the American Institute for Conservation's Guide to Digital Photography and Conservation Documentation.

## White balance

Once the workspace has been color balanced, the next step is to set the white balance for each camera. Different camera manufacturers allow this to be accomplished in different ways, so it's best to consult your user's manual for details. General guidelines are as follows:

- *Standard copy stand illumination*: One of the simplest methods is to use a gray card. This is an inexpensive flat card printed with an 18% neutral gray on the front and 80% white on the back. With the copy stand lights on and adjusted to the correct brightness, place the card gray side-up on the copy stand, fill the camera frame with card, and shoot with the camera in white balance mode. This will create a custom white balance setting for that camera and that particular lighting setup. Depending on your camera's options, it is usually possible to use the same process to manually measure the color temperature of a single light source.

- *Transmitted light*: The simplest way to color balance a light box setup is to create a custom white balance setting that matches the light source. Instead of using a grey card, one can simply use the light source as the target. The light source should have even illumination and a ground glass top such that the light is dispersed and there is a color neutral and even illuminated surface.

The image processor should keep an eye out for pictures with noticeably warm or cold color casts. These images are a good indicator that the camera white balance or lighting setup needs adjustment.

If bulbs are replaced or any other changes are made to the lighting setup or the workstations or the workspace, cameras should be re-white balanced.

## Image Capture

At the moment the standard for storing archival-quality digital images is the uncompressed TIFF. This file format contains a large amount of data in a stable format that is predicted to be supported long into the future.

A standard high-resolution workflow is to shoot in RAW mode, which prevents the camera from doing any in-camera processing to the image. Images can go directly to a tethered computer workstation or to in-camera SD cards. Depending on circumstances, a record of the objects shot can either be kept in a logbook, spreadsheet, or database automatically synced to the camera's actions. One program that can assist with the capture of images on multiple cameras is digiCamControl.

When a session has finished, the photographer then processes the RAW images using programs such as Adobe Bridge and Photoshop. The following conversion settings are recommended before basic adjustments (exposure, contrast, etc.) are made:

- Color space: Adobe (1998). This provides a larger range of color than sRGB and is more universally supported than other color spaces.

- Color depth: 16-bits per channel. A 16-bit image takes up twice as much storage space than its 8-bit counterpart but is recommended when the objects being photographed show large ranges of color.

- Output size: Best practice is to use the size native to the camera used – avoid extrapolating image data.

- Resolution: Values in the range of 600ppi are acceptable.

The image is further adjusted in Photoshop (rotated, cropped, etc.) and two copies – one of the processed file and one of original unprocessed image – are saved as TIFF files onto the processor's computer. Filenames are assigned sequentially using rendition numbers, which allow the photographer to assign as many images to a single object without resulting to awkward filenames.

Metadata, including object inventory number, a brief description, information about the institution, and a copyright tag, is then added to the files via Bridge.

If needed, the processed TIFF files can then be batch-processed via a Photoshop action into lower-resolution JPEGS for web use. Recommended JPEG parameters for screen use are a 50% reduction in physical size from the parent TIFF, a step down to 8-bit color, and a shift to 72ppi.

## Metadata

Metadata is extremely useful for identifying and tracking archival images. These XMP sidecar files can be imbedded directly into the image and stay attached unless forcibly removed. At a bare minimum, image metadata should include the name of the project, the object inventory or other identifying number, and, if warranted, a copyright tag containing the phrase "© [date], Institution Name".

Metadata templates are easily created, edited, and applied in programs such as Adobe Bridge, Photoshop, and Lightroom.

## Archival Data Storage

The *3-2-1 Rule* is a rule-of-thumb used to describe the current best practices for archiving digital image collections. It refers to the fact that there should always be at least *three* identical copies of each image, stored on *two* different types of media, with *one* copy always stored off-site on a remote server or storage system.

Even with the best equipment, digital archives can be lost or damaged without proper precautions. The most common dangers are hardware failure and user error (i.e., accidental erasure or dropped drives). Hardware failure is by far the most common threat. According to many sources, 3.5" disk drives have an average expected lifespan of only 3-5 years. These drives can last longer if not continuously powered, though it becomes more difficult to monitor drive and data health if they are stored off.

To combat this – keeping the 3-2-1 Rule in mind – a digital image archive should be stored on a working "live" hard-drive system that is backed up daily to a similar set-up (preferably in a different physical location). This data should be frequently backed up to an off-line read-only product such as DVDs or Blu-ray discs. At this time, the archival community does not recommend cloud storage as a long-term storage option for digital images, though the astronomical community has been moving in this direction for some time.

Current recommended hardware setups are:

- POD (*Pile of Drives):* Fills the two of the three stipulations of the 3-2-1 rule. Can be a drain on a computer's operating system as each drive needs its own unique connection. Advantages: cheap, easy to add capacity. Disadvantages: can be awkward to manage and maintain as a permanent archive.

- JBOD (*Just a Bunch of Disks*): A collection of hard drives in a secure enclosure that simplifies computer-to-drive connections and allows the combination of several drives into one volume (concatenation). Concatenated drives are often referred to as SPAN sets. Data can be backed up on drives within the same enclosure or to independent drives in another enclosure via Ethernet. Advantages: cuts down on the amount of electrical and data connections, easy to maintain. Drawbacks: no automatic mechanism to repair drive errors, though this can be organized via third-party software.

- RAID (Redundant Array of Independent Disks): A collection of drives configured as one volume. Data is written to multiple drives with redundancy. Mirrored (RAID 1) or double parity (RAID 6) setups are a good choice for storing working files in critical situations, such as projects with hard deadlines. Network Attached Storage (NAS) drives have become quite popular and reduce the knowledge needed to set up an array for a large amount of storage. Advantages: "self-healing," fast data access and saving. Disadvantages: complex to setup and manage. Note that a *single* RAID unit cannot be seen as a backup system as any changes made to the root data are quickly made to the copies.

Read-only options are:

- Optical storage media (including Blu-ray, DVDs, and CDs) and tape drives. These forms of media do not suffer from hardware or software problems and, because they are not rewriteable, are immune from user error such as accidental deletion.

Physical media can suffer from age-related decay ("bit-rot"), though this can be mitigated by using high-quality recording media. As of this writing Delkin Gold and Taiyo Yuden gold foil discs are recommended. Burned discs should be stored in jewel cases, or in archival-quality plastic sleeves stored in notebooks. Discs should never be marked with anything other than archival solvent-free marking pens.

## Digital Photography Resources for Archives

1. *Dpbestflow.org* (Library of Congress and ASMP clearinghouse for standards and best practice).

2. Bigras, Choquette, and Powell. *Lighting Methods for Photographing Museum Objects*, Canadian Conservation Institute, 2010.

3. Hunter, Fil. *Light: Science and Magic: An Introduction to Photographic Lighting*, 2007.

4. Krough, Peter. *The DAM Book: Digital Asset Management for Photographers*, 2009.

5. *https://cloudharmony.com/status-1year-group-by-regions-and-provider* (a comparison of the stability of different cloud storage services.)

### Notes on Contributors

Samantha van Gerbig is exhibition designer and object photographer at the Collection of Historical Scientific Instruments at Harvard University. She holds a M.Sc. in the History of Science from the University of Oxford and a B.A. from Harvard University in Visual and Environmental Studies. Some of her recent photographic work can be found in *Tangible Things: Making History through Objects* (OUP, 2015).

David Sliski is a graduate student in the Department of Physics and Astronomy at the University of Pennsylvania. From 2011 to 2014, he was part of the DASCH (Digital Access to a Sky Century @ Harvard) team.

# Preservation Recommendations for Historic Photographic Jackets

**Sara Schechner**
Harvard University, USA

**David Sliski**
University of Pennsylvania, USA

Historic photographic plates are commonly stored in paper sleeves or jackets. These enclosures serve two purposes: to protect the photographic plate from damage caused by dirt and scratches, and to record data associated with that particular image. A common issue facing astronomical glass plate collections is the deterioration of the jackets that house the plates. This deterioration threatens the integrity of the emulsion and annotations on the glass plates, as well as the loss of data from the sleeves themselves.

There are several reasons why deterioration can occur. Most jackets made before 1980 consist of wood pulp paper, which has shorter fibers than cotton and linen rag paper and so is more brittle. The production process often leaves lignin in the pulp and adds sizing for a better writing surface; both make the paper acidic and breaks down the fibers over time. The inherent acidity of wood paper is compounded by the conditions in which it is stored. Environmental pollutants such as sulfur and nitrogen oxides in the air, or acids from wood shelves and crates, are readily absorbed by paper, especially in the presence of moisture. Considering that most observatory photographic plates have been stored in wood-paper sleeves in wooden crates in non-controlled environments, serious concern is warranted. A recent study at the Harvard College Observatory, for example, found that a plate jacket, which showed visible signs of deterioration, had a pH of 3.6. This is similar to the acidity of orange juice, not something one would want protecting unique astronomical data or historically important annotations.

Although funding for the conservation of photographic plates is limited, the authors think it is imperative that collections of astronomical photographic plates consider rehousing the plates that are in jeopardy. According to the International Standards Organization, the paper enclosure should be acid-free, lignin-free, buffered material of a neutral pH between 7.0 to 9.5 ±0.2.[1] The jacket material should also pass the Photographic Activity Test. Developed by the Image Permanence Institute of the United States, this test looks at the chemical interactions that can occur between photographic images and their storage containers. If any component of an enclosure (such as inks, labels, adhesives, tapes, or paper) causes discoloration of photographic material, that product should not be placed near archival photographs.

When rehousing the photographic plates in new sleeves, care must be taken not to lose the information inscribed on the old ones. Not only is the content of these manuscripts worth preserving, but also the handwriting. The old jacket should be scanned or photographed, and its image should be printed on the new archival sleeve. We recommend that a multifunction laser printer with direct-feed capability be used to scan the old jacket and print directly to the new one. One should ensure that the toner in the printer passes the Photographic Activity Test.[2]

## Notes on Contributors

Sara J. Schechner is the David P. Wheatland Curator of the Collection of Historical Scientific Instruments, Harvard University. She is a founding member of the American Astronomical Society's Working Group for the Preservation of Astronomical Heritage and is active in similar groups worldwide. Her latest book is *Tangible Things: Making History through Objects* (OUP, 2015).

David Sliski is a graduate student in the Department of Physics and Astronomy at the University of Pennsylvania. From 2011 to 2014, he was part of the DASCH (Digital Access to a Sky Century @ Harvard) team.

## Notes

**Captions with Thumbnails, followed by Plate Images at High Resolution**
*Schechner, Sliski paper for JHA 2016*

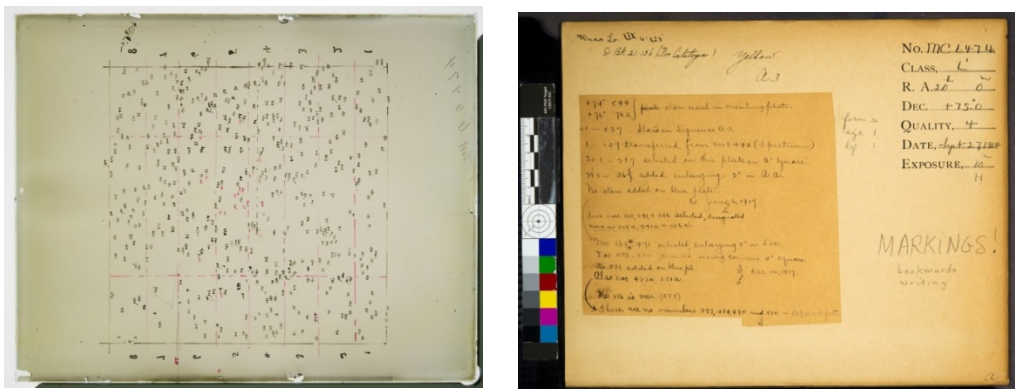

Fig. 1. Plate MC6474 and its jacket illustrate the process of compiling a catalog of stars from multiple plates that were sensitive to different portions of the spectrum.   This yellow-sensitive plate was exposed on September 27, 1914 and compared to a blue-sensitive plate.  As the work progressed through the end of 1917, different women at the observatory annotated the jacket.   They include "L" (Evelyn Leland), "S," and "AJC" (Annie Jump Cannon).  The plate annotations are unusual in being written backwards on the plain side in order to be read directly from the emulsion side.   Photographic plate taken with the 16-inch Metcalf telescope in Cambridge, MA, with its protective sleeve. Courtesy of the Harvard College Observatory Plate Stacks.

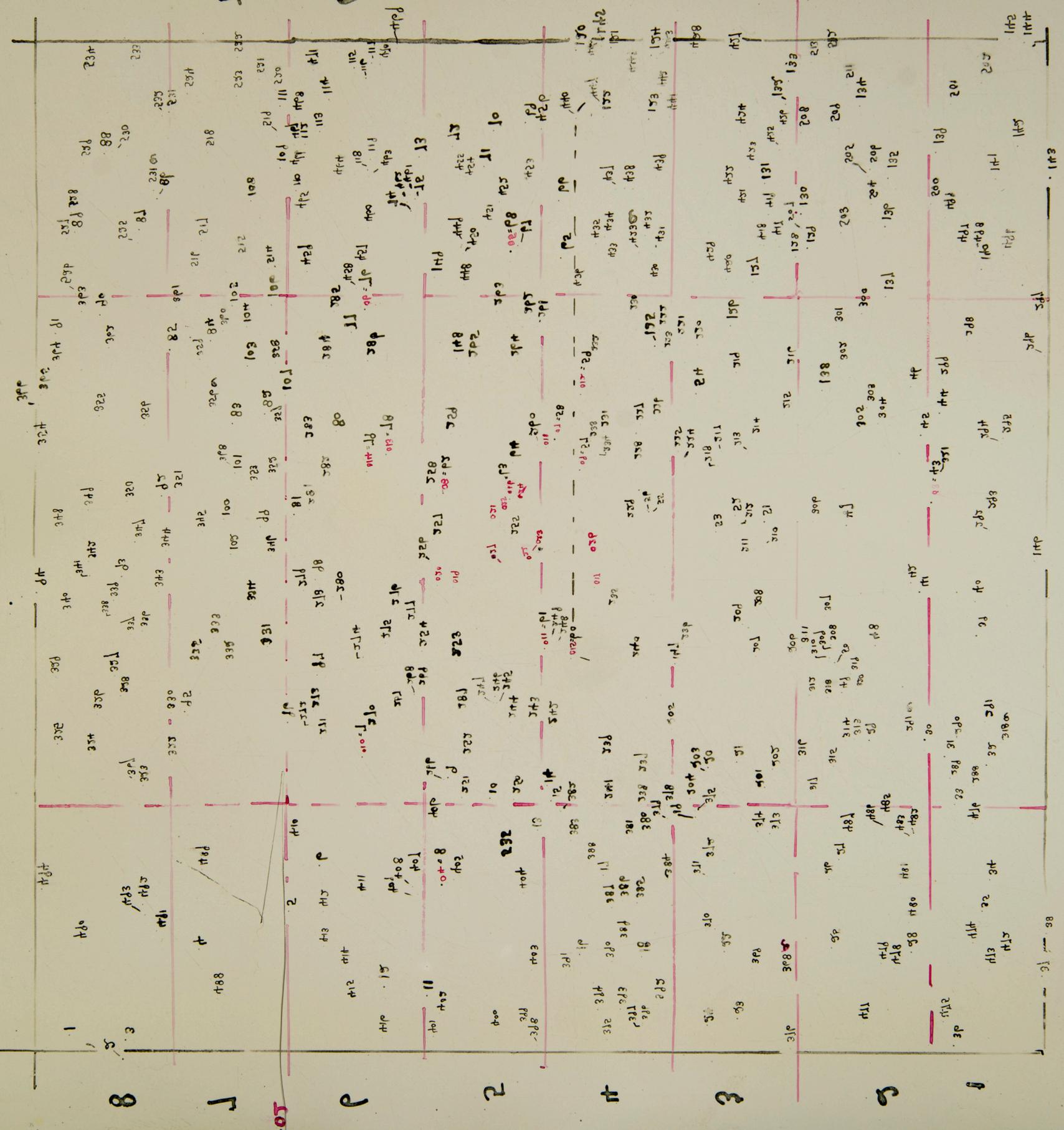

Yellow

A 3

+74° 844 } plate stars used in orienting plate.
+76° 762 }

01 — 027 . Stars in Sequence A 3.

1 — 149 transferred from Mc b 443 (Spectrum)

201 — 287 selected on this plate in 2° Square.

288 — 367 added, enlarging 2° in R.A.

No stars added on blue plate.

𝒟 June 6, 1917.

(Two nos. 318, 391, + 356 selected, designated
now as 318a, 391a + 356a.

Nos 368a — 471 selected, enlarging 2° in Dec.

Nos 472 — 534 selected using corners 4° Square.

No. 535 added on blue pl.          𝒮 Dec 10, 1917.

Also nos. 433a, 531a.          𝒮.

No. 116 is var. (ATC)

There are no numbers 377, 484, 495 and 510. — 509 is a defect.
          𝒮.

form 2
size 1
def. 1

No. MC 6474

CLASS,          L'

R.A. 20$^h$          0$^m$

DEC.          +75°.0

QUALITY,          4

DATE, Sept. 27 1914

EXPOSURE,          10$^m$
          H

MARKINGS!

backwards
writing

a

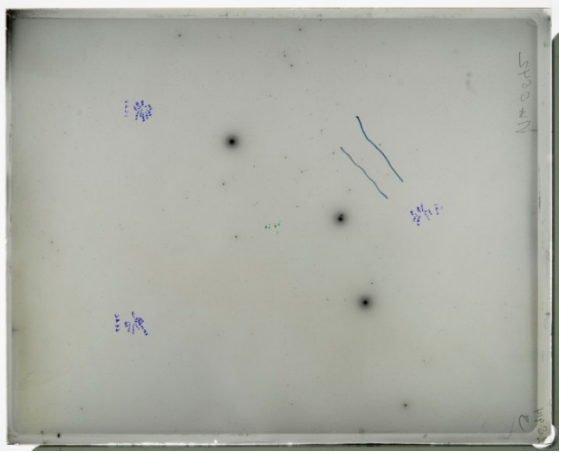

Fig. 2.   A meteor streaked across the sky when plate RH8842 was exposed on 16-17 May 1939.  Fred L. Whipple took note of it on May 26, drawing lines on either side of its visible path on the plate.  Without these annotations, it would be hard to locate the meteor on the plate, much less know that any astronomer took an interest in it.  Photographic plate RH8842 was taken with a 3-inch Ross-Fecker patrol telescope at Oak Ridge Observatory, Harvard, MA.  Courtesy of the Harvard College Observatory Plate Stacks.

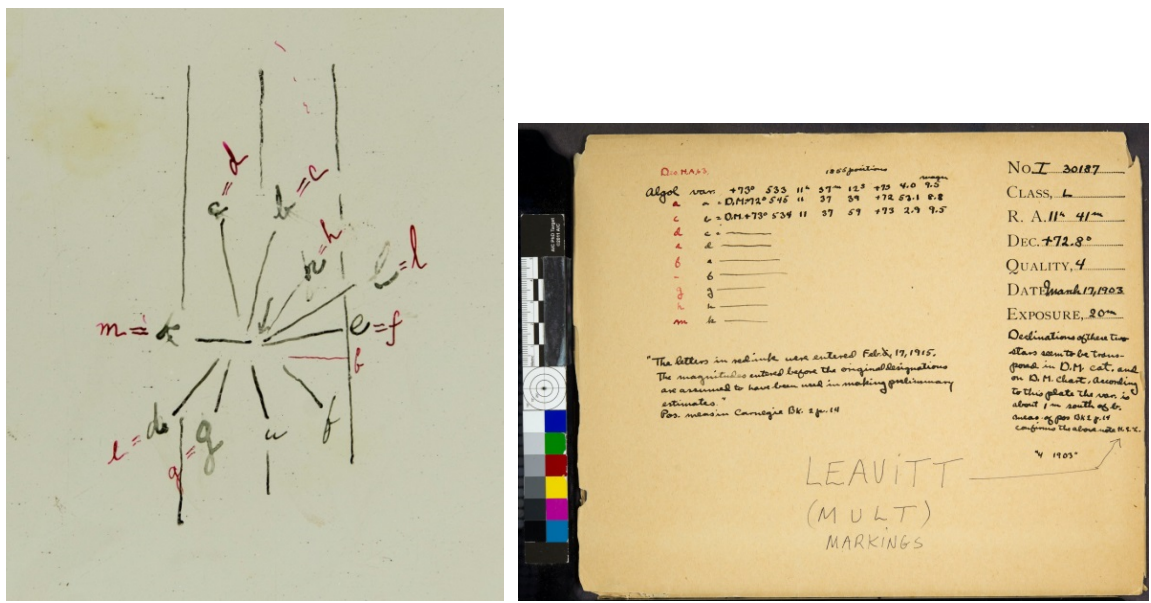

Fig. 3. Exposed on 17 March 1903, plate I30187 and its jacket illustrate Henrietta Swan Leavitt's method for discovering variable stars.  She marked the star of interest-- Zeta Draconis --with an arrow, and labeled nearby stars with letters *a* through *k* in black ink so that their magnitudes could be compared.  Zeta Draconis proved to be an Algol-type eclipsing binary star.   After Henry Norris Russell and Harlow Shapley used the light curve of the star to determine the orbital parameters of the binary system in 1914, Leavitt returned to the plate to reassign letters in red ink to the comparison stars.  Photographic plate taken with the 8-inch Draper telescope in Cambridge, MA (detail) with protective sleeve.  Courtesy of the Harvard College Observatory Plate Stacks.

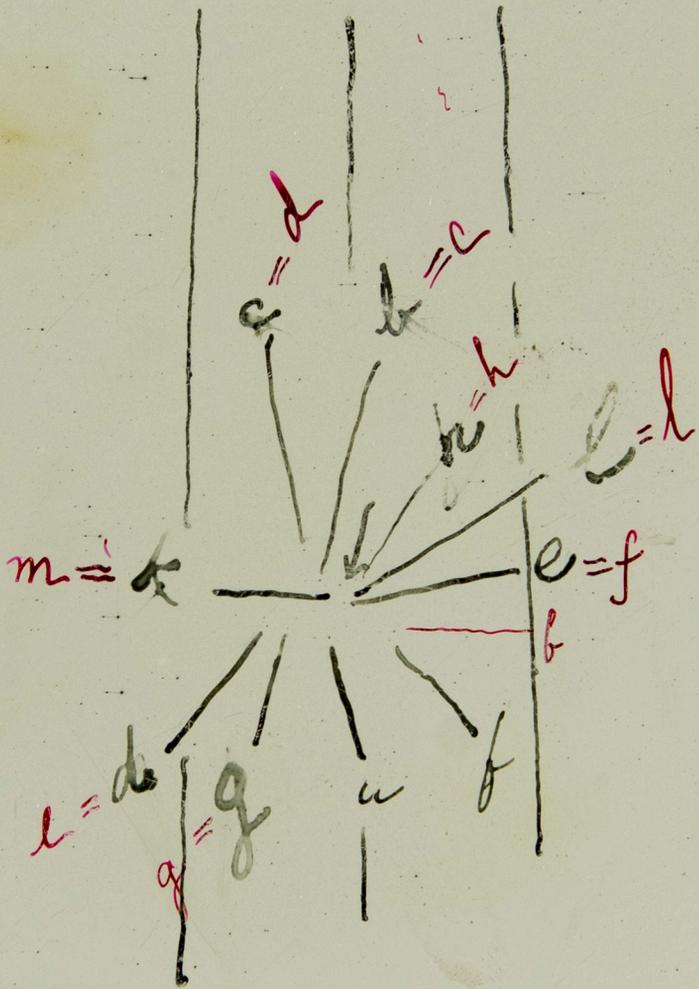

Des H.A. 63.                              1855 positions

Algol  var.  +73° 533  11ʰ 37ᵐ 12ˢ  +73 4.0  mag 9.5
  a    a = D.M. 72° 545  11  37  39  +72 53.1  8.8
  c    c = D.M +73° 534  11  37  59  +73 2.9  9.5
  d    c =
  e    d ______________
  b    e ______________
  —    b ______________
  g    g ______________
  h    h ______________
  m    k ______________

"The letters in red ink were entered Feb. 17, 1915.
The magnitudes entered before the original designations
are assumed to have been used in making preliminary
estimates."
Pos. meas in Carnegie Bk. 2 p. 14

LEAVITT ——————

(M U L T)

MARKINGS

Declinations of these two
stars seem to be trans-
posed in D.M. cat. and
on D.M. chart. According
to this plate the var. is
about 1 m south of b.
meas. of pos. Bk 2 p. 14
confirms the above note H.S.L.

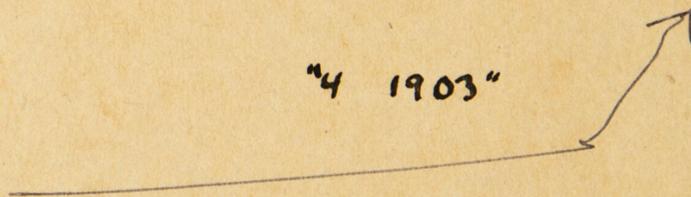
"4 1903"

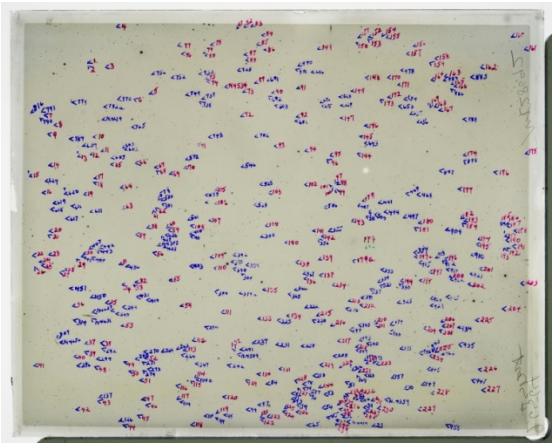

Fig. 4.  Annotations in diverse colors of ink mark the 'fuzzy' objects that could be galaxies of different sorts.  Although the photographic plate MC28092 was taken on 25-26 January 1936 with Harvard's 16-inch Metcalf telescope at Oak Ridge Observatory, its jacket indicates that the plate was later examined for Seyfert galaxies, which were not described until 1943.  Harlow Shapley and his team used the 16-inch Metcalf in conjunction with the 24-inch Bruce telescope in the southern hemisphere to identify more than 500,000 new galaxies.  Courtesy of the Harvard College Observatory Plate Stacks.

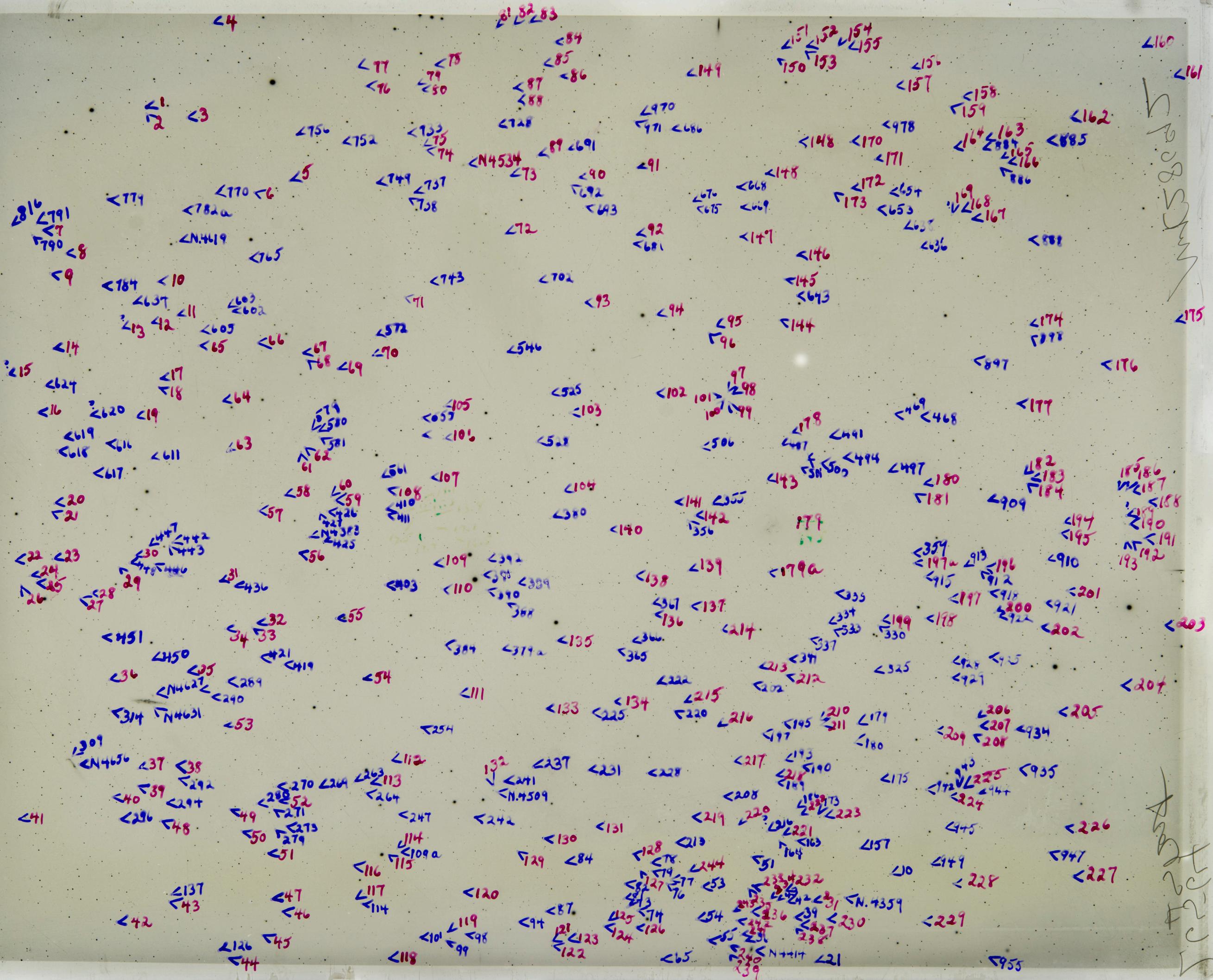

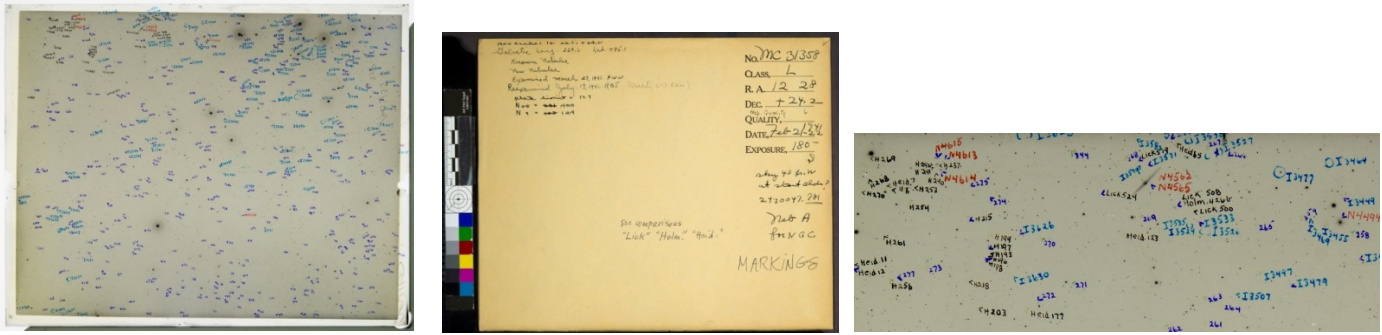

 Fig. 5.  Marked in five colors of ink by at least three individuals, plate MC31358, shown here with its protective jacket, is an example of layered annotations representing work over several years.  Taken on 21-22 February 1941, the plate was studied a month later by Frances Woodworth Wright who marked the 'fuzzy' objects on it with green arrows and two shades of blue ink.  These included Index Catalogue (IC) objects.  On July 17 another observer, "RBJ", reexamined the plate and likely added the labels in red ink for New General Catalogue (NGC) objects.  The detail of the upper portion of the plate is also filled with notes in black ink, which designate objects according to deep sky surveys by Erik Holmberg and the Lick and Heidelberg observatories.   These may be the work of a third observer after 1950.  Galaxies such as NGC 4565, a tilted spiral galaxy (seen near the upper center of the detail) were studied in the 1930s-1950s in order to determine their shapes, how they rotated, and their dynamical interactions with other galaxies. Photographic plate MC31358 taken with the 16-inch Metcalf telescope at Oak Ridge Observatory (full plate and detail) with its jacket.  Courtesy of the Harvard College Observatory Plate Stacks.

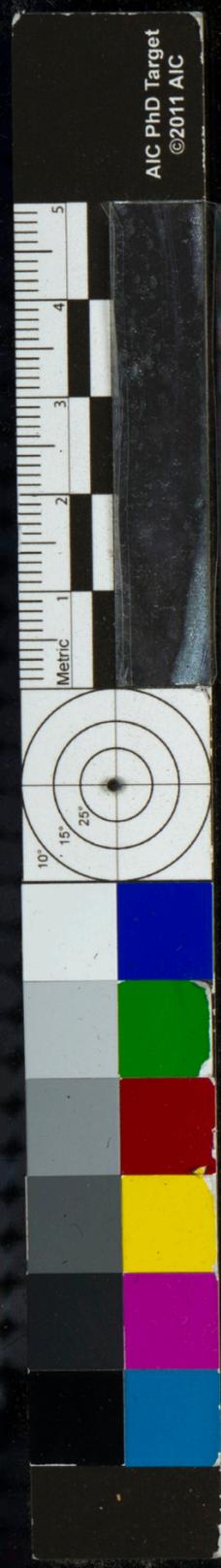

1900 center: 12 26.1, +24.2
Galactic Long. 229.6   Lat. +85.1

Known Nebulae
New Nebulae
Examined March 27, 1941. FWW
Reexamined July 17, 1941. RBJ   (Quality 6→7 FWW)

plate limit = 17.7
N₂₅ = ~~394~~ 400
N₉ = ~~122~~ 124

See comparisons
"Lick" "Holm." "Heid."

No. MC 3/358
CLASS, L
R.A. 12  28
DEC. + 24.2
Neb. Quality   6
QUALITY,
DATE, Feb 21-22 1941
EXPOSURE, 180⁻
                  S
sky 45 fr. w
at start eld?
243004?. ?81

Neb A
for NGC

MARKINGS

( ) I3623

N4615

‹H269

N4613

H242‹  ‹H237
H241

‹H268   ‹Heid.7   H240  N4614  ‹275
Heid.7      ‹H8 ‹H253
‹H270

H254

‹H215

H261

‹Heid.11
Heid.12   ‹277   273

H256

‹H218

344

I358   I355  ( ) I3558
268 ‹Lick519  ‹Heid65  267 I3527
I3571  ( 4 ) 2206
I3579?  ‹I3571

N4562

N4565
‹Lick524       Lick 508
‹Holm. H266
‹Lick 500
269
Heid.123   I3535   ‹I3533
I3524  ( ) I3526

274

I3626
270

I3630
271
‹222

‹H203   Heid 177

OI3464

I3477

‹I3449

59
I3469  I3455  258

‹I3

I3497   ‹I3479

263  ‹I3507
264

262   261

265

N4494

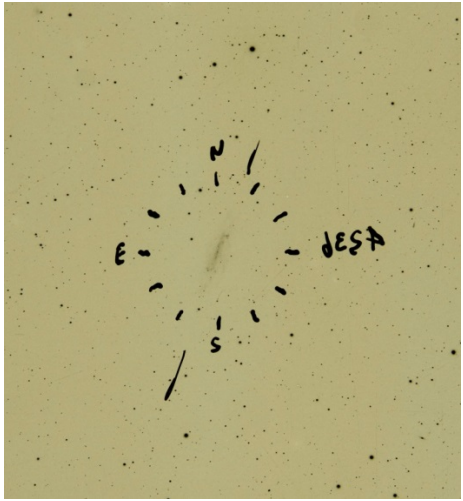

Fig. 6. The markings on plate MA8246, exposed on April 6-7, 1940, draw attention to NGC 4236 and its orientation in the sky. Edwin Hubble had classified this nebula as a Sc Spiral Galaxy in his famous 1926 paper, which divided extra-galactic nebula into a sequence of "island universes" based on their structure. At the time, Harlow Shapley had opposed the idea of galaxies beyond the Milky Way, but this 1940 photographic plate offers evidence of Harvard College Observatory's continued interest in galaxy distribution under Shapley's leadership. Photographic plate taken with the 12-inch Metcalf telescope at Oak Ridge Observatory, Harvard, MA (detail). Courtesy of the Harvard College Observatory Plate Stacks.

N

E

S

dESA

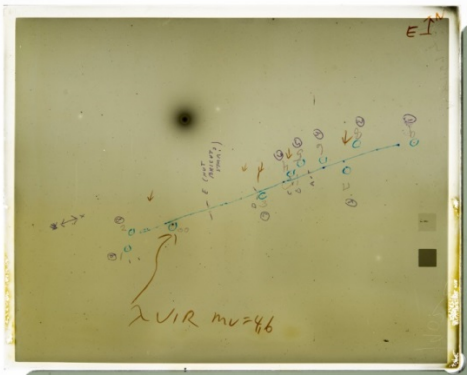

Fig. 7.  An observer on 8 April 1920 has marked plate MC16749 along the ecliptic in an effort to track a solar system object, most likely a recently discovered minor planet.   The bright object in the field is Saturn.  Photographic plate MC16749 exposed with the 16-inch Metcalf telescope in Cambridge, MA.  Courtesy of the Harvard College Observatory Plate Stacks.

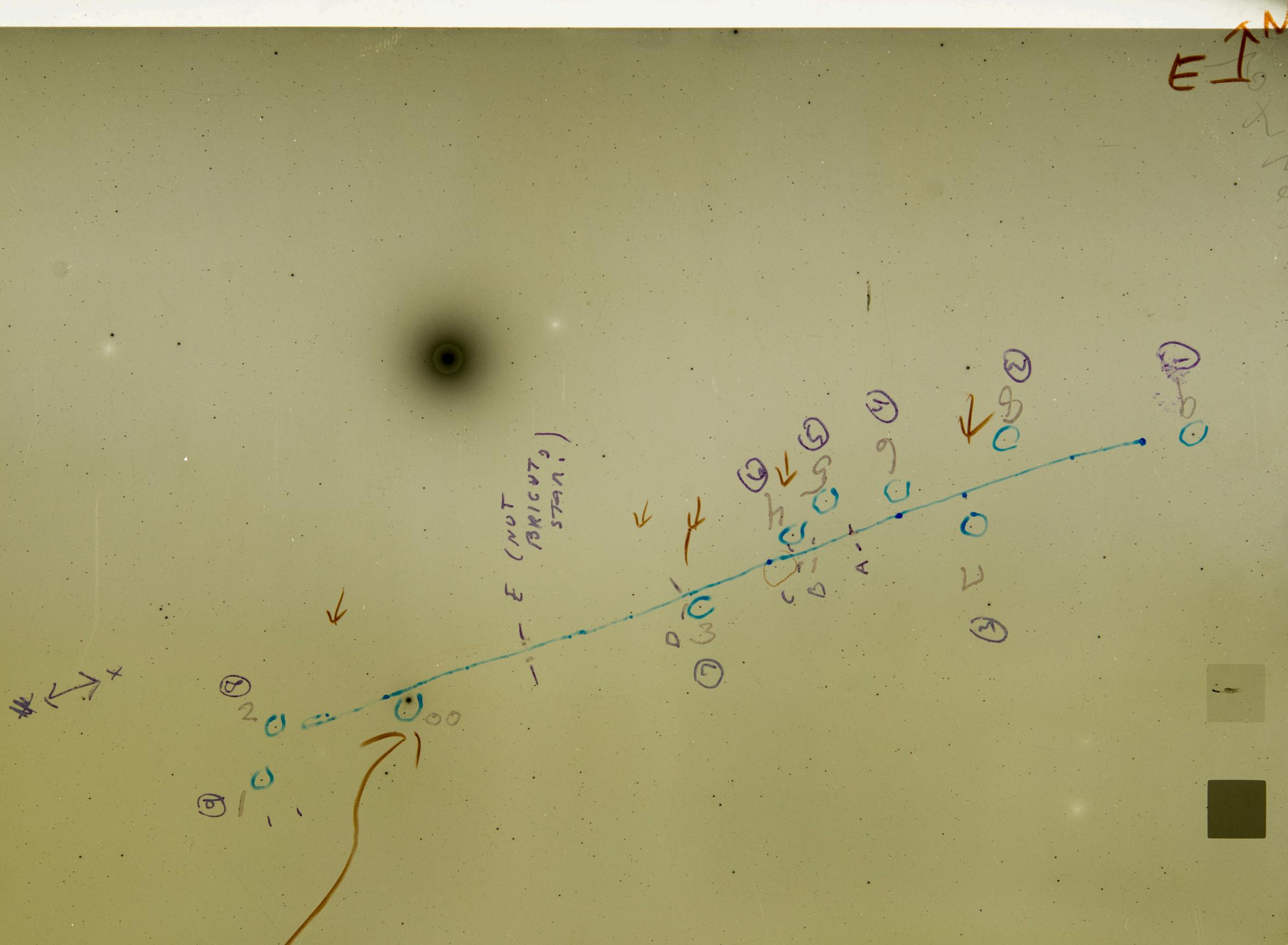

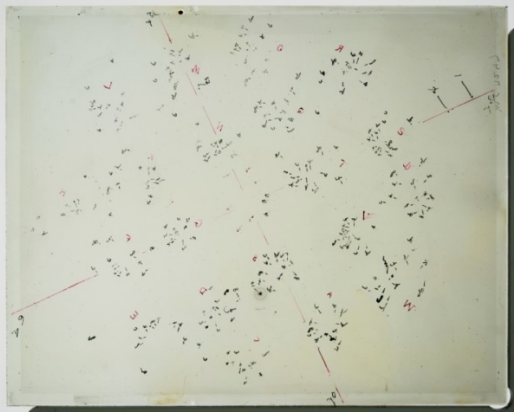

Fig. 8.  E. C. Pickering's new method of using a North Polar Sequence as a standard for photometric work is exemplified by markings on plate MC7247.   For this project, he enlisted Henrietta Swan Leavitt to determine the magnitudes of a sequence of stars photographed near the North Pole.  When stars of unknown brightness were later photographed on the same plate as stars in the North Polar Sequence, their magnitudes could be determined.  Photographic plates with a North Polar Sequence also enabled the astronomers to rate the quality of the night for photometric work.   Photographic plate MC7247 taken with the 16-inch Metcalf telescope in Cambridge, MA on 16 December 1914. Courtesy of the Harvard College Observatory Plate Stacks.

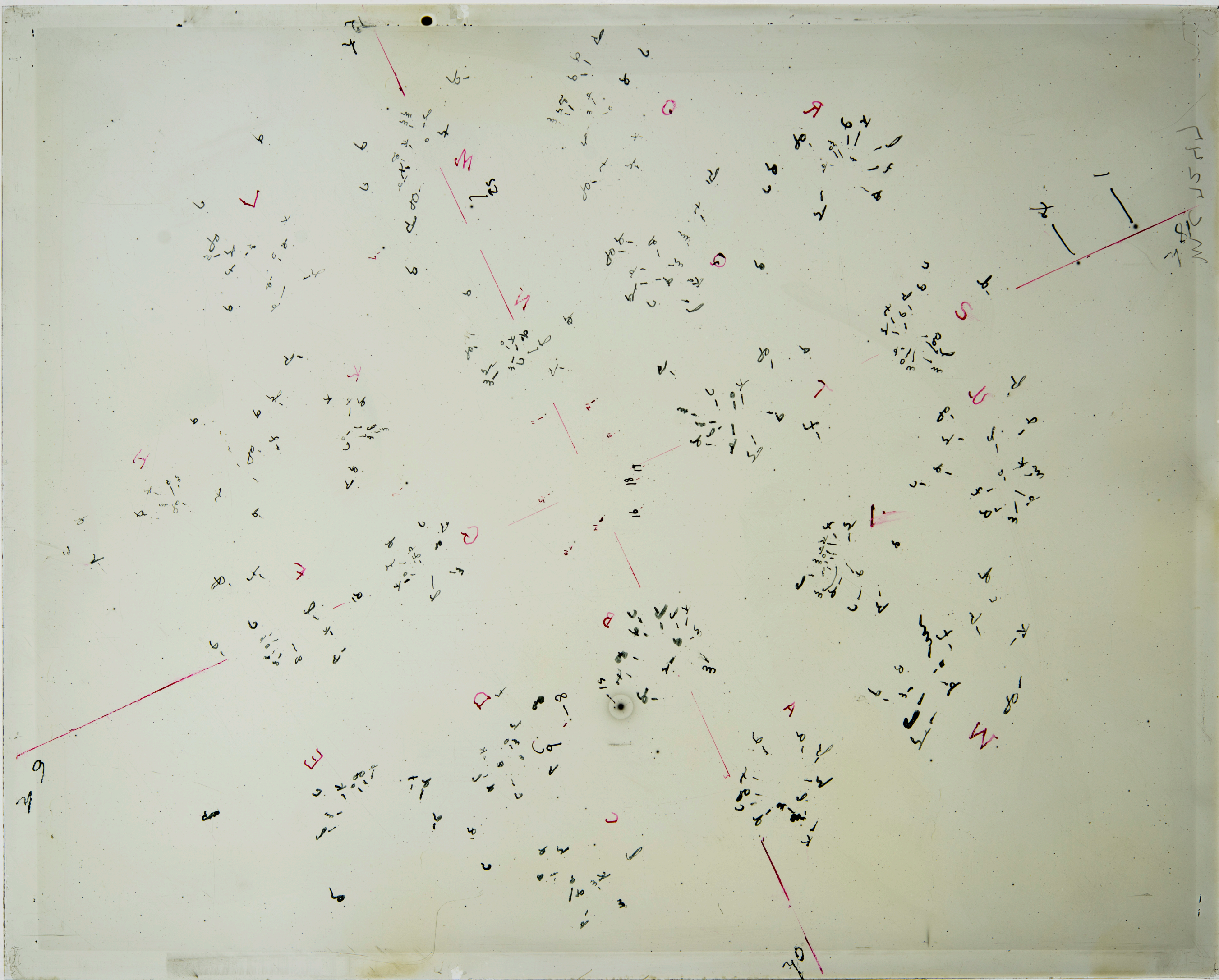